\documentclass[11pt,tightenlines,eqsecnum,floats,aps,amsmath,amssymb,nofootinbib,prd,shownopacs,floatfix]{revtex4}

\pdfoutput=1

\usepackage{amsfonts}
\usepackage{amsmath}
\usepackage{amssymb}
\usepackage{graphicx,color}
\usepackage{float}
\usepackage{hyperref}
\usepackage{subfigure}
\usepackage{dcolumn}
\usepackage{soul}
\usepackage{ulem}
\usepackage{verbatim}
\usepackage{graphicx}
\usepackage{color}
\usepackage{xparse}

\begin{document}

\title{Anisotropic non-singular quantum bounce as a seesaw \\ and amplification mechanism for magnetic fields}

\author{Meysam Motaharfar$^1$}
\email{mmotah4@lsu.edu}
 \author{Parampreet Singh$^{1,2}$}
\email{psingh@lsu.edu}
\affiliation{$^1$ Department of Physics and Astronomy, 
Louisiana State University,
  Baton Rouge, LA 70803, USA}
  \affiliation{$^2$ Center for Computation and Technology,
Louisiana State University, Baton Rouge, LA 70803, USA}

\begin{abstract}

We investigate the evolution of a homogeneous magnetic field within Bianchi-I loop quantum cosmology, in which the big bang singularity is replaced with an anisotropic quantum bounce. Using effective spacetime description, we conduct extensive numerical simulations with randomized initial conditions for two cases: first in the presence of a pure homogeneous magnetic field and then adding a massless scalar field that effectively captures the bounce regime even in the presence of inflationary potentials. For a cigar-like approach to the classical big bang, which is far more prevalent than a point-like approach, the quantum geometric bounce acts as a seesaw mechanism for magnetic field energy density. Due to the seesaw mechanism, the magnetic field energy density gets amplified by several orders of magnitude across the bounce for a small pre-bounce value of magnetic field energy density,  or vice versa. In the presence of the scalar field, the seesaw mechanism is completely replaced by amplification of the magnetic field across the bounce when the initial energy density of the scalar field significantly dominates. When the universe has a point-like approach to the classical big bang, the seesaw mechanism is absent, and magnetic field energy density experiences amplification or suppression across the non-singular bounce depending on initial densities. 

\end{abstract}

\maketitle

\section{Introduction}

Magnetic fields pervade the universe on observable scales, from planets and stars to galaxies, clusters of galaxies, and even the intergalactic medium (see, e.g., \cite{Kronberg:1993vk, Grasso:2000wj, Widrow:2002ud, Giovannini:2003yn, Kandus:2010nw, Widrow:2012, Yamazaki:2012pg, Durrer:2013pga, Subramanian:2015lua, Vachaspati:2020blt, Subramanian:2019jyd}). The observed magnetic fields in the galaxies and clusters turn out to be around a few micro-Gauss \cite{Clarke:2000bz}, while there exists a lower bound on their strength around femto-Gauss in the intergalactic medium \cite{Tavecchio:2010, Finke:2015ona, Neronov:2010gir, Taylor:2011}. The origin of such large scale correlated magnetic fields is a long-lasting problem in modern cosmology. Standard proposals to explain the origin of these magnetic fields can broadly be classified into two categories: astrophysical origin \cite{Kulsrud:2007an, Ryu:2008hi, Subramanian}, where observed magnetic fields originate from an initial weak magnetic field produced in astrophysical sources due to local mechanisms, which are then amplified and transferred from local sources to larger scales; and primordial origin \cite{Grasso:2000wj, Kandus:2010nw, Yamazaki:2012pg, Subramanian:2015lua, Subramanian:2019jyd}, where a seed magnetic field is produced in the early universe preceding the structure formation. Although astrophysical proposals such as the dynamo mechanism \cite{Brandenburg:2004jv,Subramanian:2019jyd} can amplify the amplitude of the magnetic fields in galaxies, it remains unclear how efficiently magnetic fields seed can spread in such scenarios. Therefore, understanding the primordial origin of magnetic fields on large scales is an important question. 

Several primordial mechanisms have been constructed to generate magnetic fields in the early universe, among which the magnetic fields produced during the cosmological phase transition \cite{Vachaspati:1991nm, Sigl:1996dm, Dolgov:2001nv, Stevens:2007ep, Kahniashvili:2009qi} can have relevant strength but small correlation lengths \cite{ Durrer:2003ja}, while magnetogenesis during the inflationary phase turns out to have relevant strength and correlation lengths \cite{Turner:1987bw, Ratra:1991bn, Martin:2007ue, Demozzi:2009fu, Kanno:2009ei, Sharma:2017eps, Campanelli:2013mea, Bamba:2006ga, Ferreira:2013sqa, Ferreira:2014hma, Pal:2023vti}. Unless conformal invariance is broken, the electromagnetic (EM) fields do not get amplified during the inflationary phase in FLRW spacetime. This can mainly be achieved by coupling the EM fields to higher order curvature terms or to the inflaton field. However, it turns out that in the former, the strength of the generated magnetic field is not enough, while in the latter, the mechanism suffers from either a strong coupling problem \cite{Demozzi:2009fu}, a back-reaction problem \cite{Kanno:2009ei, Sharma:2017eps}, or both (see Refs. \cite{Ferreira:2013sqa, Ferreira:2014hma, Pal:2023vti} for inflationary magnetogenesis without a strong coupling problem). One possible solution to this issue is to assume that the primordial seed magnetic field originates from a pre-big bang magnetogenesis mechanism in the bouncing scenarios \cite{Salim:2006nw, Membiela:2013cea, Sriramkumar:2015yza, Qian:2016lbf, Koley:2016jdw, Frion:2020bxc}. However, these proposals have certain limitations. First, they are not in anisotropic spacetime, which is a natural choice for spacetime in the presence of matter content with a non-vanishing anisotropic stress tensor like magnetic fields, and second, these proposals do not take into account non-perturbative quantum gravitational effects on the evolution of the magnetic field during the bounce. The main goal of this manuscript is to overcome these two limitations to understand the evolution of magnetic fields, which are produced by pre-big bang magnetogenesis, during an anisotropic quantum bounce due to quantum gravitational effects. 

One arena to understand the impact of quantum gravitational effects on the evolution of magnetic fields during the bounce is Loop Quantum Cosmology (LQC), which is a non-perturbative quantization of the homogeneous universe using techniques from Loop Quantum Gravity (LQG) \cite{Ashtekar:2011ni}. LQC predicts that the quantum gravitational effects mitigate the divergence of curvature in the Planck regime, replacing the big bang singularity with a quantum bounce leading into a non-singular spacetime \cite{Ashtekar:2006rx, Ashtekar:2006uz, Ashtekar:2006wn}. Such a prediction was first formulated in a mathematically rigorous way for a spatially flat, isotropic, and homogeneous universe where a massless scalar field was utilized as an internal clock, and it was analytically found that there exists an upper bound on the eigenvalue of the energy density operator determined by the area gap in the quantum geometry \cite{Ashtekar:2007em}. Furthermore, the probability for the occurrence of the bounce turns out to be unity in the consistent histories formulation of quantum mechanics \cite{Craig:2013mga}, while this is not the case in Wheeler-DeWitt quantum cosmology \cite{Craig:2010vf}. Quantum evolution is governed by the quantum Hamiltonian constraint, which is a second order discrete quantum difference equation resulting from underlying quantum geometry. Conducting exhaustive numerical simulations of the quantum difference equation, it was demonstrated for sharply peaked states that the bounce occurs at a fixed maximum energy density in the Planck regime, as predicted by an exactly solvable model. In fact, if the universe is macroscopic and peaked on a classical GR trajectory at late times, it remains sharply peaked across the quantum bounce \cite{Corichi:2007am}. These results of singularity resolution have been generalized to various spacetimes, including spatial curvature \cite{Szulc:2006ep, Ashtekar:2006es,Vandersloot:2006ws, Szulc:2007uk}, cosmological constant \cite{Bentivegna:2008bg, Pawlowski:2011zf, Kaminski:2009pc}, inflationary potential \cite{Giesel:2020raf}, anisotropic spacetimes \cite{Ashtekar:2009vc,Ashtekar:2009um,Wilson-Ewing:2010lkm} and in the presence of Fock quantized inhomogenities \cite{Garay:2010sk}. The underlying quantum evolution can be accurately captured for a wide class of semi-classical states using effective dynamics \cite{Taveras:2008ke}. Numerical simulations for an exactly solvable LQC model confirm the accuracy of the effective description and the occurrence of the bounce \cite{Diener:2013uka, Diener:2014mia, Diener:2014hba, Diener:2017lde}. Assuming the validity of effective dynamics, it has been shown that strong singularities are generically absent in isotropic \cite{Corichi:2008zb, Singh:2010qa} as well as anisotropic spacetimes \cite{Singh:2011gp, Corichi:2009pp, Saini:2017ipg, Saini:2017ggt, Saini:2016vgo} and there are bounds on physical observables such as energy density, expansion, and shear scalar in different models \cite{Corichi:2008zb,Gupt:2011jh, Singh:2013ava, McNamara:2022dmf}.

The magnetic fields are inherently anisotropic (with non-vanishing anisotropic stress), and for this reason, it is important to understand their generation and evolution in anisotropic spacetimes. In this manuscript, we consider loop quantized Bianchi-I spacetime, which in the isotropic limit results in spatially-flat FLRW spacetime. An important physical observable in Bianchi-I spacetime is anisotropic shear, which measures the deviation from isotropic spacetime. The anisotropic shear is a constant of motion in classical Bianchi-I spacetime
for vacuum spacetime and for matter content with a vanishing anisotropic stress tensor. 
Loop quantization of the Bianchi-I spacetime results in extending the universe to the pre-big bang branch \cite{Chiou:2006qq, Chiou:2007sp, Martin-Benito:2008dfr, Martin-Benito:2009xaf, Ashtekar:2009vc, Diener:2017lde}. Although the anisotropic shear varies during the bounce phase due to quantum gravitational effects, it is preserved in the classical regime for matter content with a vanishing anisotropic stress. More precisely, starting from a classical contracting universe evolving towards a classical expanding universe, one can show the preservation of anisotropic shear by comparing the pre-bounce and post-bounce values of the anisotropic shear at the same volume. This has been investigated in Bianchi-I LQC models, and it has been demonstrated that there is only one unique loop quantization, the Ashtekar-Wilson-Ewing quantization \cite{Chiou:2007sp, Ashtekar:2009vc}, for Bianchi-I spacetime \cite{Motaharfar:2023hil}, in which the anisotropic shear is preserved across the bounce in the low curvature regime. By contrast, magnetic fields have a non-vanishing anisotropic stress tensor, and the anisotropic shear is not conserved across the bounce in the classical regime. Moreover, the evolution of magnetic field energy density is related to the evolution of anisotropic shear and vice versa due to the non-vanishing anisotropic stress tensor. Consequently, it is expected that the magnetic field energy density has a non-trivial and rich evolution during the bounce due to the non-linear behavior of anisotropic shear and quantum gravitational effects. The relevant question is under what conditions and by what order of magnitude the magnetic field gets amplified or suppressed across the bounce in this cosmological model. We note that the evolution of magnetic fields in Bianchi-I LQC was earlier performed in Ref. \cite{Maartens:2008dd}. Our work builds upon this early work and, using extensive numerical simulations, aims to shed insights on the primordial generation of magnetic fields.

To comprehensively investigate the evolution of anisotropic shear and magnetic field across the bounce, we perform extensive numerical simulations where the universe starts in a classical regime, contracts, bounces, and expands into a classical regime. We first consider only a homogeneous magnetic field, where the universe tends to be, in general, cigar-like (two directional scale factors contract or expand, and the other one expands or contracts) rather than point-like (all three directional scale factors contract or expand). More precisely, initiating from a classical contracting universe and randomizing two directional connections, we find that the Hamiltonian constraint fixes the third directional connection such that the universe is cigar-like in approximately $93\%$ of simulations and is point-like for the rest of simulations. Comparing the pre-bounce and post-bounce values of the magnetic field energy density, we then discover that the post-bounce value of the magnetic field energy density gets suppressed or amplified across the bounce in comparison with its pre-bounce value, depending on the strength of the pre-bounce value of the magnetic field energy density. In fact, the bounce acts as a seesaw mechanism for magnetic field energy density, whereby the post-bounce value of magnetic field energy density gets amplified in comparison with its pre-bounce value up to seven orders of magnitude in some cases for small pre-bounce values of magnetic field energy density and vice versa, starting from a classical contracting cigar-like universe. In other words, the post-bounce value of magnetic field energy density gets suppressed in comparison with its pre-bounce value for all considered initial conditions with a pre-bounce value of magnetic field energy density approximately larger than $10^{-6}$ (in Planck units) and amplified for all considered initial conditions with a pre-bounce value of magnetic field energy density approximately less than $10^{-11}$ (in Planck units), starting from a classical contracting cigar-like universe. By varying the value of triads, or correspondingly, the volume, we find that the cutoff for which the post-bounce value of magnetic field energy density gets suppressed across the bounce in comparison with its pre-bounce value for all considered initial conditions is independent of the volume. For a classical contracting point-like universe, we find that the ratio of post-bounce to pre-bounce of the magnetic field energy density can either be smaller or greater than unity, depending on the pre-bounce value of the magnetic field energy density. While we do not observe the seesaw mechanism in this case, we find that the post-bounce value of magnetic field energy density gets suppressed in comparison with its pre-bounce value for all considered initial conditions when the pre-bounce value of magnetic field energy density is approximately larger than $10^{-6}$ (in Planck units), as is the case for the cigar-like universe.

To understand the impact of the presence of other matter content in the universe on the observed seesaw mechanism, we add a massless scalar field. It has been earlier shown that in the presence of inflationary potentials in Bianchi-I LQC, the kinetic term of the scalar field dominates the potential term \cite{Gupt:2013swa}. Hence, the massless scalar field, in some sense, captures the bounce regime of inflationary dynamics. Here we find that adding the massless scalar field affects the seesaw mechanism. Starting from a classical contracting cigar-like universe, we find that the post-bounce value of the magnetic field energy density always gets amplified across the bounce in comparison with its pre-bounce value when the initial energy density of the massless scalar field is approximately greater than 13 times the initial energy density of the magnetic field. Starting from a classical contracting point-like universe, we find that as the initial energy density of the massless scalar field significantly dominates, the Hamiltonian constraint is not satisfied for magnetic field energy density approximately larger than $10^{-6}$, as a result of which there is no cutoff value below which the post-bounce value of magnetic field energy density gets suppressed in comparison with its pre-bounce value for all considered initial conditions. This means that the post-bounce value of magnetic field energy density gets suppressed or amplified in comparison with its pre-bounce value, depending on the pre-bounce value of magnetic field energy density. These results indicate that the magnetic field’s evolution during the quantum bounce depends on the initial conditions (cigar or point-like universe), the strength of the magnetic field energy density, and the equation of the state of other matter content in the universe. Hence, if the primordial seed of the magnetic field is generated before the big bang, one must consider the evolution of the magnetic field during the bounce phase to match its strength to the observed value. Interestingly, this amplification occurs for a small value of the magnetic field, so if the magnetogenesis mechanism is not efficient enough to amplify the magnetic field, it is possible that it gets amplified during the bounce phase to match the observed value. Conversely, if it turns out that the magnitude of the magnetic field is significantly large, it gets suppressed and becomes inefficient in explaining observational constraints.

The manuscript is organized as follows: In Section \ref{section II}, we provide a brief review of the classical dynamics of Bianchi-I cosmology in terms of symmetry-reduced Ashtekar-Barbero variables, namely the triads and the corresponding directional connections. We then establish their relationship to the familiar metric variables, discuss the dynamics of a pure homogeneous magnetic field in classical spacetime, and show the existence of singularities by finding the general solutions. In Section \ref{section III}, we briefly review the effective dynamics of the Bianchi-I LQC model and derive Hamilton's equations. Then, we present the numerical results in Section \ref{section IV}, showing that singularities are replaced by a quantum bounce for a universe filled with a homogeneous magnetic field, and discuss the evolution of the magnetic field across the bounce, starting from either the classical cigar-like or point-like universe in the contracting branch. We also present a similar analysis for a universe filled with a homogeneous magnetic field along with a massless scalar field while varying the initial ratio of the massless scalar field to the magnetic field energy density. It should be noted that we use Planck units for numerical simulations throughout the manuscript. Finally, we give the summary and conclusions in Section \ref{section V}.

\section{Classical dynamics of Bianchi-I spacetime with a magnetic field}\label{section II}

The inclusion of cosmological magnetic field breaks the isotropy of spacetime; hence, it is natural to consider it in anisotropic cosmology, with one of the simplest settings being the Bianchi-I spacetime, whose metric is given by
\begin{align}\label{Bianchi-I}
\mathrm{d}s^2 = - \mathrm{d}t^2 + a_{1}^2 \mathrm{d}x^2 + a_{2}^2 \mathrm{d}y^2 + a_{3}^2 \mathrm{d}z^2 .
\end{align}
Here we have set the lapse function to be unity, i.e., $N=1$, and $a_{1}, a_{2}$, and $a_{3}$ are directional scale factors with the mean scale factor defined as $a:=(a_{1}a_{2}a_{3})^{1/3}$. In the isotropic limit, the metric (\ref{Bianchi-I}) reduces to the Friedmann-Lemaître-Robertson-Walker (FLRW) metric, describing a spatially flat, homogeneous, and isotropic universe. The spatial topology can be non-compact $(\mathbb{R}^3)$ as well as compact $(\mathbb{T}^3)$, and the choice does not matter for our analysis as the Ashtekar-Wilson-Ewing loop quantization prescription used in the next section is unaffected by this choice \cite{Corichi:2009pp}. If the spatial topology is non-compact, one needs to introduce a fiducial cell to define a symplectic structure. We set the coordinate volume of this cell to be unity.

The loop quantization program is based on the classical gravitational phase space variables, namely the Ashtekar-Barbero connections $A_{a}^{i}$ and the triads $E_{i}^{a}$ (where $i=1, 2, 3$), so it is useful to briefly review the dynamics of Bianchi-I spacetime in a canonical framework and then relate it to the conventional metric variables. The Ashtekar-Barbero variables $A^{i}_{a}$ and $E_{i}^{a}$ reduce to connections $c^{i}$ and triads $p_{i}$ with only one independent component per spatial direction upon symmetry reduction and imposing the Gauss and the spatial-diffeomorphism constraints. One can then show that the triads are kinematically related to the directional scale factors as follows:
\begin{align}\label{triads}
|p_{1}| = a_{2}a_{3}, \ \ \ \ \ \ \ \ \ \ \ \ \ \ \ \ \ |p_{2}| = a_{1}a_{3}, \ \ \ \ \ \ \ \ \ \ \ \ \ \ \ \ \  |p_{3}| = a_{1}a_{2}.
\end{align}
The modulus sign arises because of the orientation of the triads. However, without any loss of generality, we here assume a positive sign for the orientation of directional triads. From these relations, one can find that directional scale factors are related to triads, i.e., $a_{1} = \sqrt{p_{2}p_{3}/p_{1}}$ (and similarly, $a_{2}$ and $a_{3}$ with cyclic permutations). Moreover, in the phase space, the triads and connections satisfy the following Poisson bracket:
\begin{align}
\{c^{i}, p_{j}\} = {8\pi G} \gamma \delta^{i}_{j},
\end{align}
where $\gamma \approx 0.2375$ is the Barbero-Immirzi parameter, which is fixed by black hole thermodynamics in LQG. The classical Hamiltonian constraint for matter content minimally coupled to the gravitational sector, in terms of directional connections $c_{i}$ and triads $p_{i}$, reads as
\begin{align}\label{Hamiltonian}
\mathcal{H}_{cl} = \mathcal{H}_{g} + \mathcal{H}_{m} = - \frac{1}{8 \pi G \gamma^2 v} \left(c_{1}p_{1} c_{2}p_{2} + c_{2}p_{2}c_{3}p_{3} + c_{3}p_{3}c_{1}p_{1}\right) + \mathcal{H}_{m},
\end{align}
where $\mathcal{H}_{m} = \rho v$ is the matter part of the Hamiltonian, with $v = \sqrt{p_{1}p_{2}p_{3}} = a_{1}a_{2}a_{3}$ being the physical volume of a unit co-moving cell and $\rho$ being the energy density. Given the classical Hamiltonian $\mathcal{H}_{cl}$, the dynamical equations for triads and connection components are determined using Hamilton's equations as follows:
\begin{align} \label{Hamilton's equations}
\dot p_{i} = \{p_{i}, \mathcal{H}_{cl}\} = - 8 \pi G \gamma \frac{\partial \mathcal{H}_{cl}}{\partial c_{i}}, \ \ \ \ \ \ \ \ \ \ \ \ \dot c_{i} = \{c_{i}, \mathcal{H}_{cl}\} = 8 \pi G \gamma \frac{\partial \mathcal{H}_{cl}}{\partial p_{i}},
\end{align}
where a dot denotes the derivative with respect to the cosmic time $t$. The first set of equations leads to $c_{i} = \gamma \dot a_{i} = \gamma H_{i} a_{i} $, which together with the Hamiltonian constraint $\mathcal{H}_{cl} \approx 0$ result in the following equation for Bianchi-I spacetime
\begin{align}\label{First-Friedmann}
H_{1}H_{2}+ H_{2}H_{3} + H_{3}H_{1} &= 8 \pi G \rho .
\end{align}
%
Here $H_{i}$ denotes directional Hubble parameters, which are related to the time derivatives of triad components, such as
\begin{align}\label{directional-Hubble}
H_{1} \equiv \frac{\dot a_{1}}{a_{1}}= \frac{1}{2} \left(\frac{\dot p_{2}}{p_{2}} + \frac{\dot p_{3}}{p_{3}} - \frac{\dot p_{1}}{p_{1}}\right),
\end{align}
(and similarly for $H_{2}$ and $H_{3}$). Likewise, using the dynamical equations for directional connections, $c_{i}$, and taking the time derivative from the first set of Hamilton's equations for triads, one finds
%
\begin{align}\label{Raychaudhuri}
\dot H_{2} + \dot H_{3} + H^{2}_{2} + H_{3}^2 + H_{2}H_{3}  = \frac{8 \pi G}{v} \left(p_{2}\frac{\partial \mathcal{H}_{m}}{\partial {p}_{2}} + p_{3} \frac{\partial \mathcal{H}_{m}}{\partial {p}_{3}}\right) .
\end{align}
For matter content with a vanishing anisotropic stress tensor one gets 
\begin{align}
{p_{2}}\frac{\partial \mathcal{H}_{m}}{\partial {p}_{2}}  = {p_{3}} \frac{\partial \mathcal{H}_{m}}{\partial {p}_{3}} =  \frac{v}{2} \frac{\partial \mathcal{H}_{m}}{\partial v} = - \frac{P}{2} v,
\end{align}
with $P$ defined as the isotropic pressure, i.e., $P = -{\partial \mathcal{H}_{m}}/{\partial v}$. Then Eq. (\ref{Raychaudhuri}) reduces to 
\begin{align}\label{second-Friedmann}
     \dot H_{2} + \dot H_{3} + H^{2}_{2} + H_{3}^2 + H_{2}H_{3} & = - 8 \pi G P .
\end{align}
Note that in the isotropic limit,  $H_{1} = H_{2} = H_{3} = H$ with $H = (H_{1} + H_{2} + H_{3})/3$ being the mean Hubble parameter, and Eqs. (\ref{First-Friedmann}) and (\ref{second-Friedmann}) reduce into first and second Friedmann equations for FLRW metric given by
\begin{align}
    H^{2} = \frac{8\pi G}{3}\rho, \ \ \ \ \ \ \ \ \ \ \ \ \dot H = - 4\pi G (\rho + P).
\end{align}
However, given Eqs. (\ref{First-Friedmann}) and (\ref{second-Friedmann}), one can also derive a set of generalized Friedmann equations containing information about the anisotropy of spacetime 
\begin{align}\label{generalized}
H^2 = \frac{8 \pi G}{3} \rho + \frac{\sigma^2}{6}, \ \ \ \ \ \ \ \ \ \ \ \ \ \ \dot H = - 4\pi G \left(\rho + P\right) - \frac{ \sigma^2}{2} .
\end{align}
The anisotropic shear scalar $\sigma^2$ measures the deviation from isotropic spacetime. It is the traceless part of the expansion tensor, and is measured in terms of directional Hubble parameters $H_{i}$ given by
\begin{align}\label{shear-tensor}
\sigma^2 = \sigma^{\mu\nu} \sigma_{\mu\nu} = \frac{1}{3} \left(\left(H_{1}- H_{2}\right)^2 + \left(H_{2}- H_{3}\right)^2 + \left(H_{3}- H_{1}\right)^2 \right).
\end{align}
One can also use the Hamilton's equations to show that
\begin{align} \label{pij}
\frac{\mathrm{d}}{\mathrm{d}t}(p_{i}c_{i}) = 8 \pi G \gamma v^2 \left(\rho + p_{i} \frac{\partial \rho}{\partial p_{i}}\right),
\end{align}
which for matter content with vanishing anisotropic stress tensor, i.e., ${p_{2}}\frac{\partial \rho}{\partial {p}_{2}}  = {p_{3}} \frac{\partial \rho}{\partial {p}_{3}}$, yields 
\begin{align}\label{pc-cp}
\frac{\mathrm{d}}{\mathrm{d}t} \left(p_{i}c_{i} - p_{j}c_{j}\right) = 0,
\end{align}
and using $c_{i} = \gamma H_{i} a_{i}$, one can find that the anisotropic shear scalar depletes by $1/a^6$, more precisely $\sigma^2 = {\Sigma^2}/{(6 a^6)}$, with $\Sigma^2$ being a constant of the motion for matter content with vanishing anisotropic stress tensor (see, e.g., \cite{Motaharfar:2023hil} for details). From dynamical equations for matter content with a vanishing anisotropic stress, it is obvious that the $H$, $\rho$, and $\sigma^2$ go to infinity as the scale factor goes to zero, signaling the divergence of curvature invariants and the breakdown of geodesic evolution at the singularities. However, in the presence of matter content with a non-vanishing anisotropic stress tensor, like a homogeneous magnetic field, $\Sigma^2$ is not a constant anymore (since Eq. (\ref{pc-cp}) does not hold in this case).

It turns out that the conductivity of the universe can be proportional to the average temperature of the universe at early times, turning our universe into a highly conductive universe \cite{Baym:1997gq}.
In such a highly conductive universe, the electric field is effectively zero, and one can solve Maxwell equations assuming a homogeneous magnetic field in the $x$-direction without any loss of generality. Let us consider the magnetic field energy density as earlier studied in LQC \cite{Maartens:2008dd}, given by
\begin{align}\label{mag-energy-density}
\rho_{B} = \frac{1}{2} B_{\mu}B^{\mu} = \frac{\beta^2}{2 (a_{2}a_{3})^2} = \frac{\beta^2}{2 p_{1}^2} ,  \ \ \ \ \ B^{1}  = \frac{\beta}{a^3},
\end{align}
where $\beta$ is a constant. Since magnetic field is anisotropic matter content, the Friedmann Eqs. (\ref{First-Friedmann}) and (\ref{second-Friedmann}) read as 
\begin{align}
H_{1}H_{2} + H_{2}H_{3} + H_{3}H_{1} &= 8 \pi G \rho_{B}, \label{First-Friedmann-magnetic}\\
    \dot H_{2} + \dot H_{3} + H^{2}_{2} + H_{3}^2 + H_{2}H_{3} \label{second-Friedmann-magnetic} & = 8 \pi G \rho_{B},\\
    \dot H_{1} + \dot H_{3} + H^{2}_{1} + H_{3}^2 + H_{1}H_{3} \label{third-Friedmann}& = -8 \pi G \rho_{B},\\ \dot H_{1} + \dot H_{2} + H^{2}_{1} + H_{2}^2 + H_{1}H_{2} & = - 8 \pi G \rho_{B} .\label{forth-Friedmann}
\end{align}
Combining Eqs. (\ref{third-Friedmann}) and (\ref{forth-Friedmann}) with Eq. (\ref{First-Friedmann-magnetic}), after a straightforward manipulation, one finds
\begin{align}
    H_{1} + H_{I} = \frac{\gamma_{I}}{a^3},  \ \ \ \ \ \ I =2, 3, \label{magnetic-equations}
\end{align}
where $\gamma_{I}$ are constants (not to be confused with the Barbero-Immirzi parameter $\gamma$). Multiplying the left hand side of Eqs. (\ref{magnetic-equations}), one can also reach the following equation:
\begin{align}\label{FE-magnetic}
H_{1}^2 a^6  = \gamma_{2}\gamma_{3} - 4 \pi G \beta^2 a_{1}^2.
\end{align}
It turns out that
\begin{align}
    \gamma_{2}\gamma_{3}>0,  \ \ \ \ \ \ a_{1} \leq a_{1}^{\text{max}} = \frac{\gamma_{2}\gamma_{3}}{4 \pi G \beta^2}.
\end{align}
Hence, the solution for Eq. (\ref{FE-magnetic}) is given by (see details in Ref. \cite{Jacobs:1969qca})
\begin{align}
a & \propto (1+f^2)f^{\pm \alpha -1}, \ \ \ \ \ \ \ 
f(a_{1}) = \frac{a_{1}^{\text{max}}}{a_{1}}\left [1- \sqrt{1- \left(\frac{a_{1}}{a_{1}^{\text{max}}}\right)^2}\right ], \label{solution1} \ \ \ \ \ \ \
\alpha  = \frac{\gamma_{2}+\gamma_{3}}{\sqrt{\gamma_{2}+ \gamma_{3}}}, \\
a_{I} & = (1+f^2)f^{\frac{\pm \gamma_{I}}{\sqrt{\gamma_2\gamma_3}}} \label{solution2},
\end{align}
where $\pm$ denotes the expanding (contracting) branches when the mean scale factor reaches $a_{1}^{\text{max}}$. Given Eqs. (\ref{solution1}) and (\ref{solution2}), one can see that the volume goes to zero or infinity when $a_{1}$ goes to zero, depending on the value of $\alpha$. Therefore, Eqs. (\ref{solution1}) and (\ref{solution2}) indicate the existence of singularities in the presence of a homogeneous magnetic field in the classical Bianchi-I spacetime. Therefore, it is pertinent to ask whether the loop quantization of Bianchi-I spacetime in the presence of a magnetic field is non-singular. To address this question, we briefly review the effective dynamics of Bianchi-I spacetime in the next section, including a homogeneous magnetic field.

\section{Effective dynamics of Bianchi-I LQC}\label{section III}

In LQC, one employs LQG techniques to quantize cosmological spacetimes using classical gravitational phase space variables, namely Ashtekar-Barbero variables: holonomies of connections along closed loops and fluxes of triads. This formulation yields a second order discrete quantum difference equation governing the entire evolution of the universe, even deep in the Planck regime. Notably, this equation converges to the continuous Wheeler-DeWitt equation in GR in the low curvature limit. Several studies have shown that for various cosmological spacetimes, like the isotropic model \cite{Diener:2013uka, Diener:2014mia, Diener:2014hba} and the Bianchi-I model \cite{Diener:2017lde}, the underlying quantum dynamics can be achieved through a continuum effective description under reasonable assumptions for a class of semi-classical states. For instance, the effective Hamiltonian for Bianchi-I LQC can be derived by substituting classical directional connections $c_{i}$ with bounded trigonometric functions, specifically, $c_{i} \rightarrow \sin(\bar \mu_{i} c_{i})/\bar \mu_{i}$; therefore,
\begin{align}
\mathcal{H} = - \frac{1}{8 \pi G \gamma^2 v} \left(\frac{\sin (\bar \mu_{1} c_{1})}{\bar \mu_{1}} \frac{\sin (\bar \mu_{2} c_{2})}{\bar \mu_{2}} p_{1}p_{2} + \textrm{cyclic permutations}\right) + \mathcal{H}_{m},
\end{align}
where 
\begin{align}
\bar \mu_{1} = \lambda \sqrt{\frac{p_{1}}{p_{2}p_{3}}},\ \ \ \ \ \ \ \ \ \ \ \bar \mu_{2} = \lambda \sqrt{\frac{p_{2}}{p_{1}p_{3}}}, \ \ \ \ \ \ \ \ \ \ \ \bar \mu_{3} = \lambda \sqrt{\frac{p_{3}}{p_{1}p_{2}}},
\end{align}
with $\lambda^2 = \Delta = 4 \sqrt{3}\pi \gamma \l_{Pl}^2$, and $\Delta$ being the minimum eigenvalue of the area operator in LQG. While there can be in principle other choices of $\bar \mu_i$, as shown in Ref. \cite{Motaharfar:2023hil}, there is only one unique loop quantization for Bianchi-I spacetime, Ashtekar-Wilson-Ewing quantization (a generalized version of the $\bar \mu$ scheme in the standard isotropic LQC), which successfully recovers the physical properties of GR at large volumes, namely satisfying the classicality condition and preserving anisotropic shear across the bounce for matter content with vanishing anisotropic stress tensor. This result is on the lines of similar findings in the isotropic model \cite{Corichi:2008zb, Corichi:2009pp}.

Given the holonomy edge lengths, the effective Hamiltonian for $\bar\mu$ quantization can be rewritten as
\begin{align}
\mathcal{H} = - \frac{v}{8 \pi G \gamma^2 \lambda^2} \left({\sin (\bar \mu_{1} c_{1})} {\sin (\bar \mu_{2} c_{2})}+ \textrm{cyclic permutations}\right) + \mathcal{H}_{m},
\end{align}
where the matter Hamiltonian is $\mathcal{H}_{m} = \rho_{B}v$ with $\rho_{B} = \beta^2/(2 p_{1}^2)$. Given the effective Hamiltonian, one can find Hamilton's equations:
\begin{align}\label{triads}
\dot p_{1} = \frac{p_{1}}{\gamma \lambda} \cos(\bar \mu_{1} c_{1}) \left(\sin(\bar \mu_{2} c_{2}) + \sin(\bar \mu_{3}c_{3})\right) ,
\end{align}
\begin{align}\label{connections}
\nonumber \dot c_{1} &= 8 \pi G \gamma \frac{\partial \mathcal{H}_{m}}{\partial p_{1}}  -\frac{v}{2\gamma \lambda^2
p_{1}} \left[ \bar \mu_{1} c
_{1} \cos(\bar\mu_{1}c_{1}) \left(\sin(\bar \mu_{2}c_{2}) + \sin(\bar\mu_{3}c_{3})\right)  \right. \\ &  \left. \nonumber - \bar \mu_{3} c
_{3} \cos(\bar\mu_{3}c_{3}) \left(\sin(\bar \mu_{1}c_{1}) + \sin(\bar\mu_{2}c_{2})\right)  - \bar \mu_{2} c
_{2} \cos(\bar\mu_{2}c_{2}) \left(\sin(\bar \mu_{3}c_{3}) + \sin(\bar\mu_{1}c_{1})\right)  \right. \\ &  \left. + \left(\sin(\bar\mu_{1}c_{1}) \sin(\bar \mu_{2}c_{2}) + \sin(\bar\mu_{2}c_{2} ) \sin(\bar \mu_{3}c_{3}) + \sin(\bar \mu_{3}c_{3} ) \sin(\mu_{1}c_{1})\right)\right],
\end{align}
while the other four Hamilton's equations (two for triads and two for connections) can be obtained using cyclic permutation. From the Hamiltonian constraint, i.e., $\mathcal{H}\approx 0$, one can find the energy density in terms of trigonometric functions as follows:
\begin{align}
\rho = \frac{1}{8\pi G \gamma^2 \lambda^2} \left({\sin (\bar \mu_{1} c_{1})} {\sin (\bar \mu_{2} c_{2})} + {\sin (\bar \mu_{2} c_{2})} {\sin (\bar \mu_{3} c_{3})} + {\sin (\bar \mu_{3} c_{3})} {\sin (\bar \mu_{1} c_{1})}\right).
\end{align}
From the expression for energy density, one finds that it is universally bounded from above with the maximum energy density $\rho_\textrm{max}$ given by
\begin{align}
\rho \leq \rho_{\textrm{max}} = \frac{3}{8 \pi G \gamma^2 \lambda^2} \simeq 0.41 \rho_{Pl}. 
\end{align}
Then, using Hamilton's equations, the directional Hubble parameters are given by
\begin{align}
\nonumber H_{1} &= \frac{1}{2 \gamma \lambda} \left( \sin(\bar \mu_{1} c_{1}) (\cos(\bar \mu_{2}c_{2}) + \cos(\bar \mu_{3}c_{3})) + \sin(\bar \mu_{2} c_{2}) (\cos(\bar \mu_{3}c_{3}) - \cos(\bar \mu_{1}c_{1})) \right. \\ & \left. + \sin(\bar \mu_{3} c_{3}) (\cos(\bar \mu_{2}c_{2}) - \cos(\bar \mu_{1}c_{1}))\right),
\end{align}
(and similarly for $H_{2}$ and $H_{3}$) and the anisotropic shear scalar also reads as
\begin{align}\label{shear}
\nonumber \sigma^2 & = \frac{1}{3\gamma^2\lambda^2} \left[\left(\cos(\bar \mu_{1}c_{1}) (\sin( \bar \mu_{2} c_{2}) + \sin(\bar \mu_{3}c_{3})) - \cos(\bar \mu_{2}c_{2}) (\sin( \bar \mu_{3} c_{3}) + \sin(\bar \mu_{1}c_{1}))\right)^2 \right. \\  &  \nonumber  \left. +  \left(\cos(\bar \mu_{2}c_{2}) (\sin(\bar \mu_{3} c_{3}) + \sin(\bar \mu_{1}c_{1})) - \cos(\bar \mu_{3}c_{3}) (\sin(\bar \mu_{1} c_{1}) + \sin(\bar \mu_{2}c_{2}))\right)^2 \right. \\  &  \left. +  \left(\cos(\bar \mu_{3}c_{3}) (\sin(\bar \mu_{1} c_{1}) + \sin(\bar \mu_{2}c_{2})) - \cos(\bar \mu_{1}c_{1}) (\sin( \bar \mu_{2} c_{2}) + \sin(\bar \mu_{3}c_{3}))\right)^2 \right],
\end{align}
 with the maximum value  given by
\begin{align}
\sigma^2 \leq \sigma^2_{\textrm{max}} = \frac{10.125}{3\gamma^2 \lambda^2} \simeq 11.57 l_{Pl}^{-2}.
\end{align}
Similarly, one can easily show that the expansion rate is universally bounded. The boundedness of energy density, expansion rate, and anisotropic shear scalar strongly indicate the resolution of classical singularities due to quantum gravitational effects in the Bianchi-I LQC model. An interesting point about the anisotropic shear in the Bianchi-I LQC model is that although it is conserved in the classical regime, as discussed in the previous section, it is not preserved during the quantum phase, even for matter content with a vanishing anisotropic stress tensor. In fact, although Eq. (\ref{pc-cp}) still holds for matter content with vanishing anisotropic stress tensor, the classical relation between directional connections and directional Hubble parameters does not hold anymore in the quantum regime, i.e., $c_{i} \neq \gamma H_{i} a_{i}$, due to polymerization of directional connections. However, it has been shown in Ref. \cite{Motaharfar:2023hil} that the anisotropic shear is preserved across the bounce in the low curvature limit for vacuum spacetime and matter content with vanishing anisotropic stress tensor. Since a homogeneous magnetic field is inherently anisotropic and the anisotropic shear evolves during the quantum phase, the evolution of magnetic field energy density across the bounce is not clear from Hamilton's equations.

Hence, the goal of the next section is to confirm the boundedness of physical quantities using numerical simulations and show that the big bang singularity is replaced with an anisotropic quantum bounce for a pure homogeneous magnetic field and also a homogeneous magnetic field along with a massless scalar field. Then we analyze the evolution of a homogeneous magnetic field energy density across an anisotropic quantum bounce.

\section{Numerical Analysis of Effective Dynamics}\label{section IV}

As mentioned earlier, using exhaustive numerical simulations, it was recently shown in Ref. \cite{Motaharfar:2023hil} that there is a unique loop quantization (Ashtekar-Wilson-Ewing quantization) for the Bianchi-I LQC model in which the anisotropic shear is preserved across the bounce in the classical regime for vacuum spacetime and matter content with vanishing anisotropic stress tensor such as massless scalar fields, dust, and radiation fields. This agrees with classical GR since, in the absence of an anisotropic stress tensor, the anisotropic shear is a constant of motion \cite{Tsagas:2007yx}. However, since magnetic fields have a non-vanishing anisotropic stress tensor, the anisotropic shear evolves in the classical regime. Due to the complicated form of Hamilton's equations and the non-linear behavior of the anisotropic shear, the quantum gravitational effects on the evolution of the anisotropic shear and the magnetic field energy density during the bounce phase are not clear for the Bianchi-I LQC model. Hence, the goal of this section is to numerically solve Hamilton's equations to confirm the existence of the bounce in the presence of a pure homogeneous magnetic field given by \eqref{mag-energy-density}, and also a homogeneous magnetic field along with a massless scalar field in the Bianchi-I LQC, and then find the evolution of the anisotropic shear and the magnetic field energy density across the bounce. In particular, we are interested in finding under what conditions and by what order of magnitude the magnetic field energy density gets amplified or suppressed across the bounce.

\subsection{Anisotropic non-singular quantum bounce as a seesaw mechanism for magnetic fields}

\begin{figure}
    \centering
    \includegraphics[scale = 0.65]{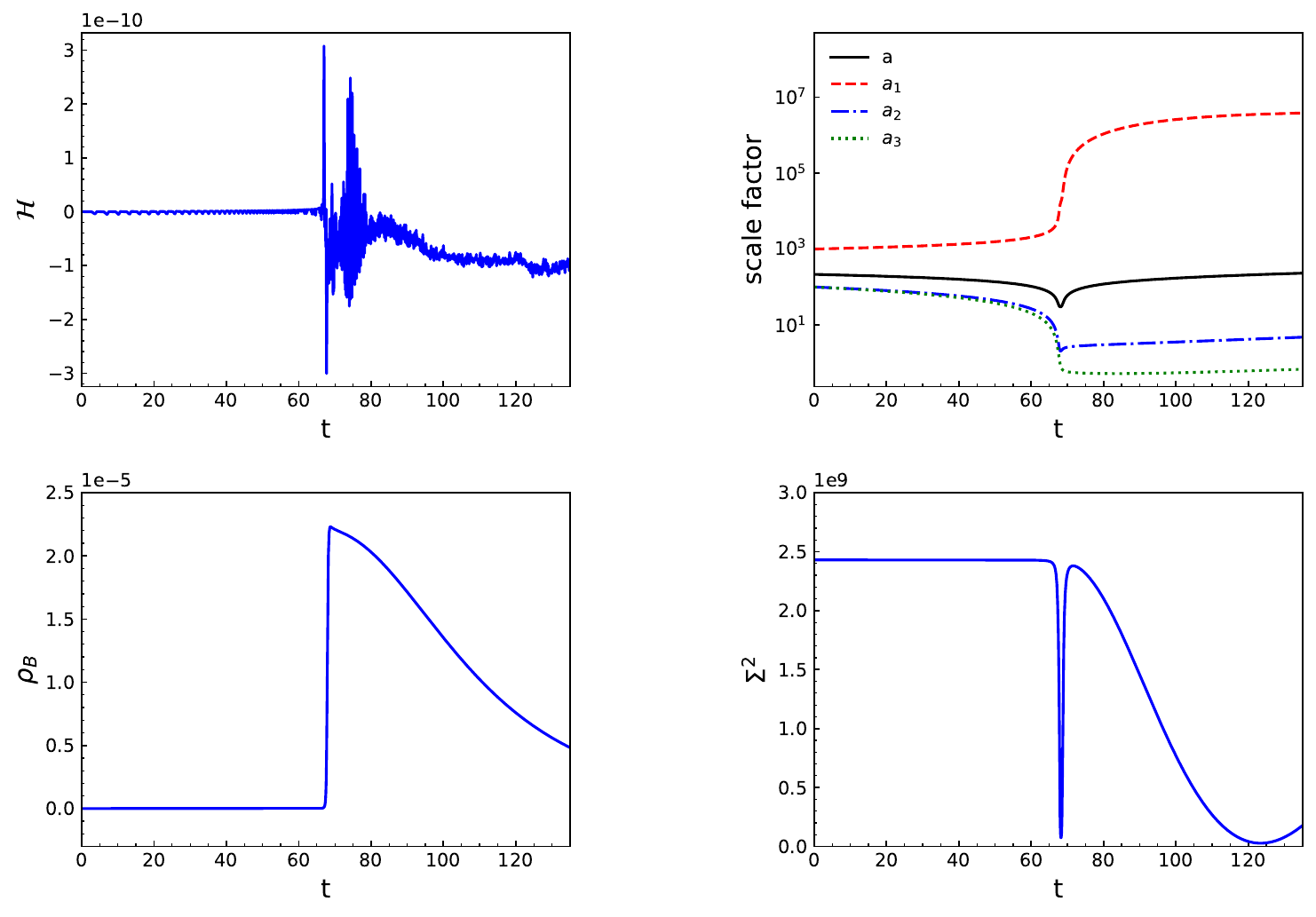}
    \caption{The time evolution of the Hamiltonian constraint (top left), directional scale factors and the mean scale factor (top right), the magnetic field energy density (bottom left), and $\Sigma^2$ (bottom right) starting from a classical contracting cigar-like evolution in the pre-bounce regime where the initial value of triads and connections is fixed at $t=0$ with $p_{1}(0) = 10^4$, $p_{2}(0) = p_{3}(0) = 10^5$, $c_{1}(0) = 1.16$, $c_{2}(0) = -0.212$, $c_{3}(0) = -0.255$, and $\rho_{B}(0) = 5\times 10^{-13}$. All physical quantities are evaluated in Planck units.}
    \label{cigar-like-scale-factor}
\end{figure}

\begin{figure}
    \centering
    \includegraphics[scale = 0.65]{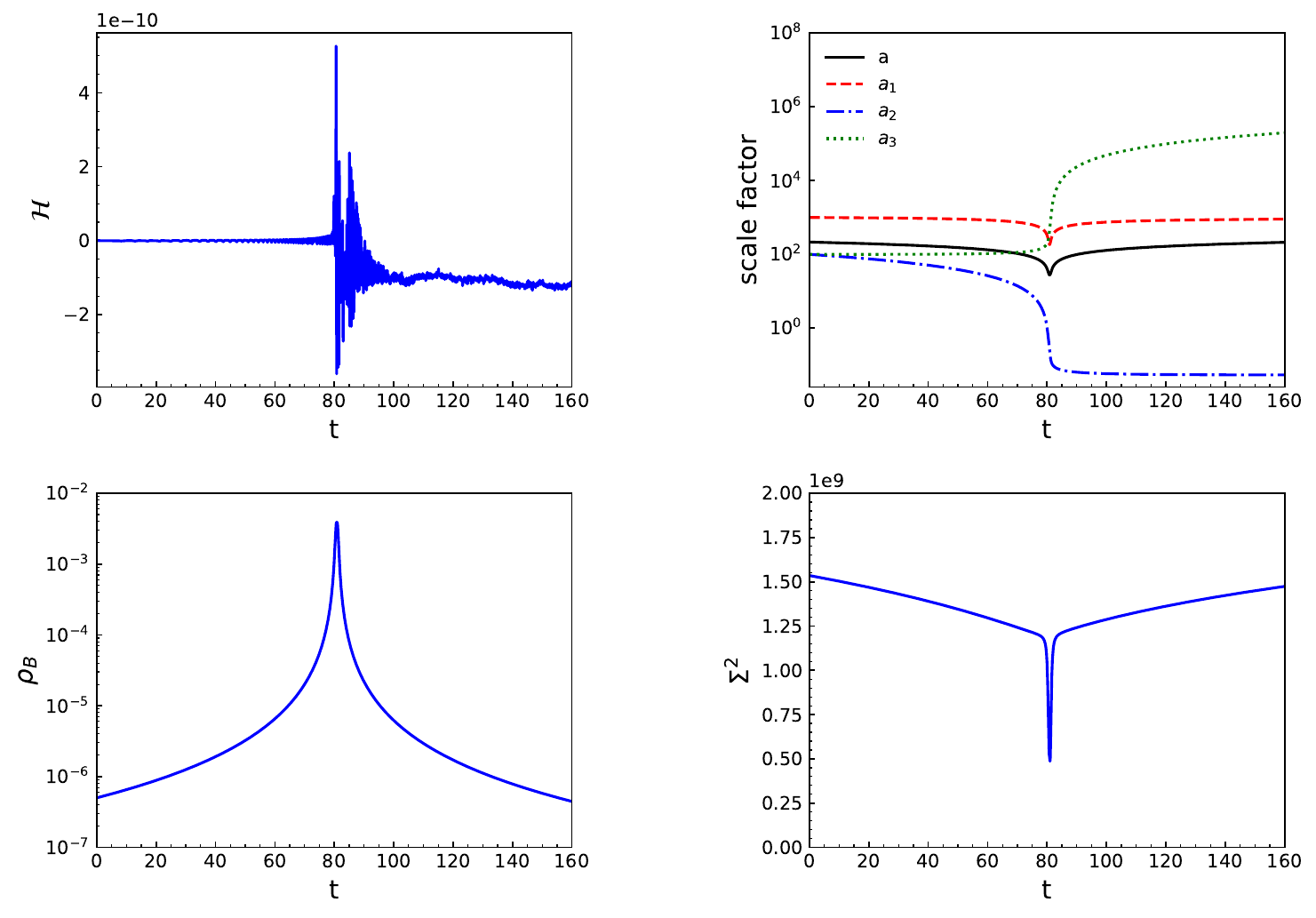}
    \caption{For a classical contracting point-like pre-bounce evolution with initial conditions set to be $p_{1}(0) = 10^4$, $p_{2}(0) = p_{3}(0) = 10^5$, $c_{1}(0) = -0.175$, $c_{2}(0) = - 0.291$, $c_{3}(0) = -0.00635$, and $\rho_{B}(0) = 5\times 10^{-7}$, the time evolution of the Hamiltonian constraint, directional scale factors and the mean scale factor, the magnetic field energy density, and $\Sigma^2$ are shown in the top left, top right, bottom left, and bottom right panels, respectively.}
    \label{point-like-scale-factor}
\end{figure}

Due to the complicated form of Hamilton's equations, one needs to numerically solve six coupled first order equations given by Eqs. (\ref{triads}) and (\ref{connections}) to fully understand the dynamical evolution of a homogeneous magnetic field in the Bianchi-I LQC model. To this end, we set the initial conditions in such a way that the universe starts in a classical regime, contracts, bounces, and expands into a classical regime. Hence, we fix all triads at $t=0$ with comparatively large values, i.e., $p_{1}(0) = 10^{4}$ (hereafter all quantities are in Planck units in this section), $p_{2}(0)=10^{5}$, and $p_{3}(0)=10^{5}$, which sets the initial volume to be $v(0) = 10^{7}$, and randomize $c_{2}$ and $c_{3}$ and finally determine the first directional connection $c_{1}$ from the Hamiltonian constraint, i.e., $\mathcal{H} \approx 0$, in the effective spacetime description of LQC, while forcing all three to be within the range that satisfy the classicality condition, i.e., $|\bar \mu_{i}c_{i}| \simeq n\pi$. We should point out that for the classical contracting point-like evolution in the pre-bounce regime, we set all three directional connections to be negative, while for the classical contracting cigar-like evolution in the pre-bounce regime, we fix $c_{1}$ to be positive and two others to be negative. Moreover, we change the magnetic field energy density uniformly by varying $\beta$, which is a constant appearing in the definition of the magnetic field energy density \eqref{mag-energy-density}. We then use the \textsc{numbalsoda} package written in Python, which is for solving ordinary differential equation initial value problems, and then use the ``\textsc{dop853}" algorithm, which is an explicit Runge-Kutta of order 8(5, 3) due to Dormand and Prince with adaptive step size control. The advantage of this algorithm is that it uses higher order corrections to adapt the step size to reduce the error propagating in the system of equations. Moreover, one can also tune the relative and absolute tolerances to fully control the numerical error. We often set the absolute and relative tolerances to be $10^{-14}$ and $10^{-11}$, respectively. To study the impact of quantum gravitational effects during the bounce phase, we compare pre-bounce and post-bounce values of physical observables at the same volume to determine the evolution of anisotropic shear and magnetic field energy density. In doing so, we use the interpolate module from the \textsc{scipy} package and use the cubic method to interpolate the post-bounce value of the anisotropic shear and the magnetic field energy density at the same volume, i.e., $v=10^7$.

Before proceeding to find out the way the pre-bounce and post-bounce values of anisotropic shear and magnetic field energy density change across the bounce, let us first understand singularity resolution. As an illustrative example, we plot the time evolution of the Hamiltonian constraint, directional scale factors, the mean scale factor, the magnetic field energy density, and $\Sigma^2$ while the initial value of triads and connections is fixed at $t=0$ with $p_{1}(0) = 10^4$, $p_{2}(0) = p_{3}(0) = 10^5$, $c_{1}(0) = 1.16$, $c_{2}(0) = -0.212$, $c_{3}(0) = -0.255$ (classical contracting cigar-like evolution in the pre-bounce regime), and $\rho_{B}(0) = 5\times 10^{-13}$ in Fig. \ref{cigar-like-scale-factor}. As it is obvious from the top right, bottom left, and bottom right panels, the mean scale factor $a$, the magnetic field energy density $\rho_{B}$, and the anisotropic shear $\Sigma^2$ do not diverge in the entire time evolution, signaling that the singularity is resolved in the Bianchi-I LQC model. Moreover, from the top left panel, one can see that the Hamiltonian constraint vanishes throughout the entire time evolution with a great accuracy. In addition, from the bottom left and bottom right panels, it is obvious that the magnetic field energy density and anisotropic shear have different values before and after the bounce, indicating that the magnetic field can either be suppressed or amplified across the bounce depending on the initial conditions. In Fig. \ref{point-like-scale-factor}, we plot the time evolution of the Hamiltonian constraint, directional scale factors, the mean scale factor, the magnetic field energy density, and $\Sigma^2$ while the initial value of triads and connections is fixed at $t=0$ with $p_{1}(0) = 10^4$, $p_{2}(0) = p_{3}(0) = 10^5$, $c_{1}(0) = -0.175$, $c_{2}(0) = - 0.291$, $c_{3}(0) = -0.00635$ (classical contracting point-like evolution in the pre-bounce regime), and $\rho_{B}(0) = 5\times 10^{-7}$. Again, one can see that all the relevant physical quantities are finite, and the singularity is resolved. Moreover, the anisotropic shear scalar and the magnetic field energy density have different pre-bounce and post-bounce values similar to the cigar-like evolution.

 \begin{figure}
    \centering
    \includegraphics[scale=0.4]{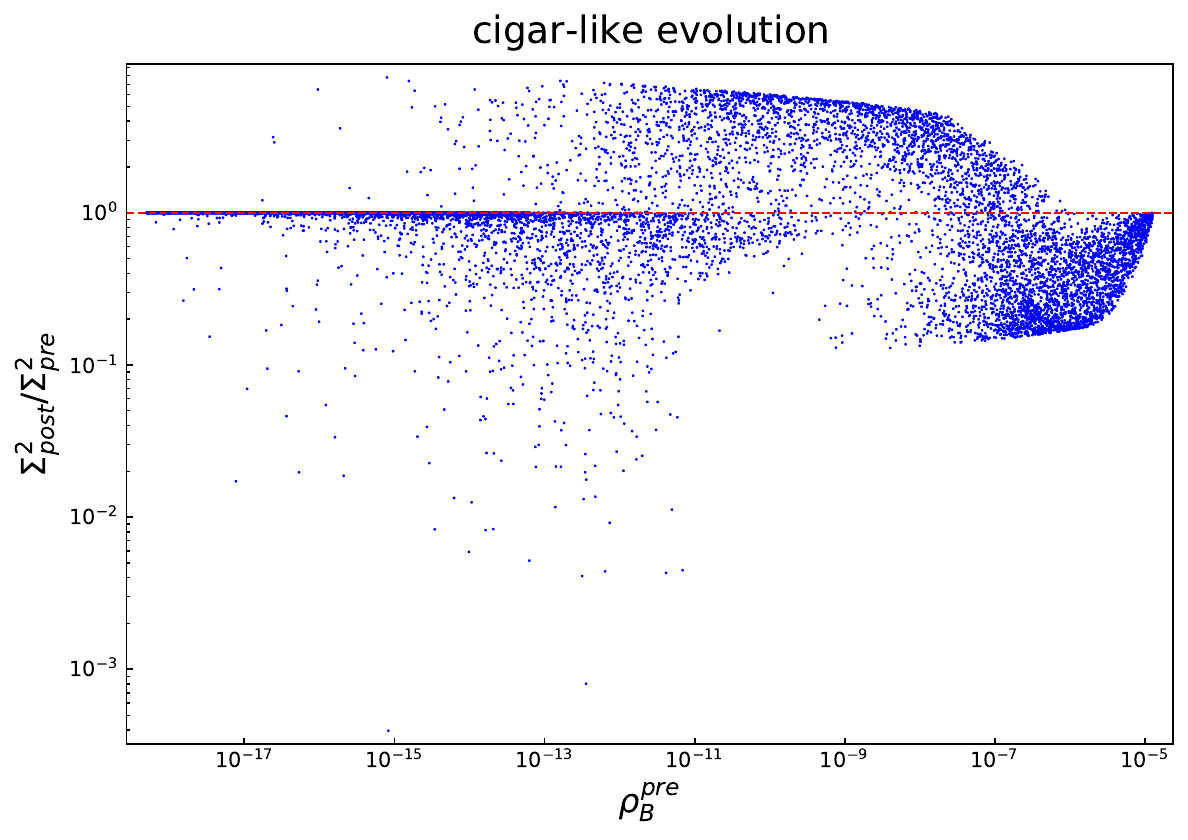}  \ \
    \includegraphics[scale = 0.4]{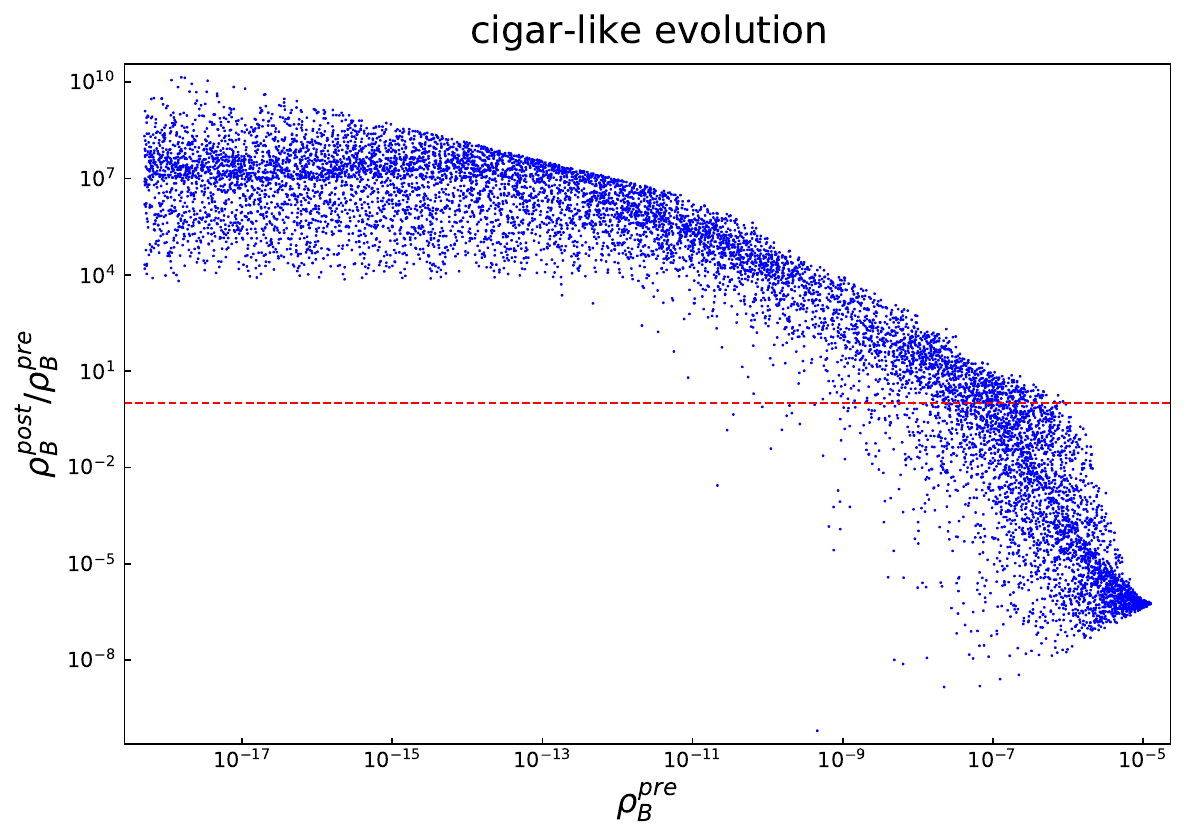}
    \caption{The ratio of post-bounce to pre-bounce value of $\Sigma^2$ versus pre-bounce value of magnetic field energy density (left) and the ratio of post-bounce to pre-bounce value of magnetic field energy density versus pre-bounce value of magnetic field energy density (right) starting from a classical cigar-like evolution in the pre-bounce regime with the initial value of the triads is fixed at $t=0$ with $p_{1}(0) = 10^{4}$, $p_{2}(0)= p_{3}(0) = 10^{5}$, and $c_{2}(0)$ and $c_{3}(0)$ are randomized, and $c_{1}(0) $ is then determined by the Hamiltonian constraint within the range that satisfy the classicality condition, i.e., $|\bar \mu_{i} c_{i}|<0.01$. It should be noted that hereafter each plot for the ratio of post-bounce to pre-bounce values of anisotropic shear and magnetic field energy density versus pre-bounce values of magnetic field energy density contains more than 10000 simulations, and the post-bounce value of physical quantities is compared to the corresponding pre-bounce value at the initial volume, i.e., $v = 10^{7}$. The red dashed line is for the ratio to be unity.}
    \label{fig1}
\end{figure}

\begin{figure}
    \centering
    \includegraphics[scale=0.4]{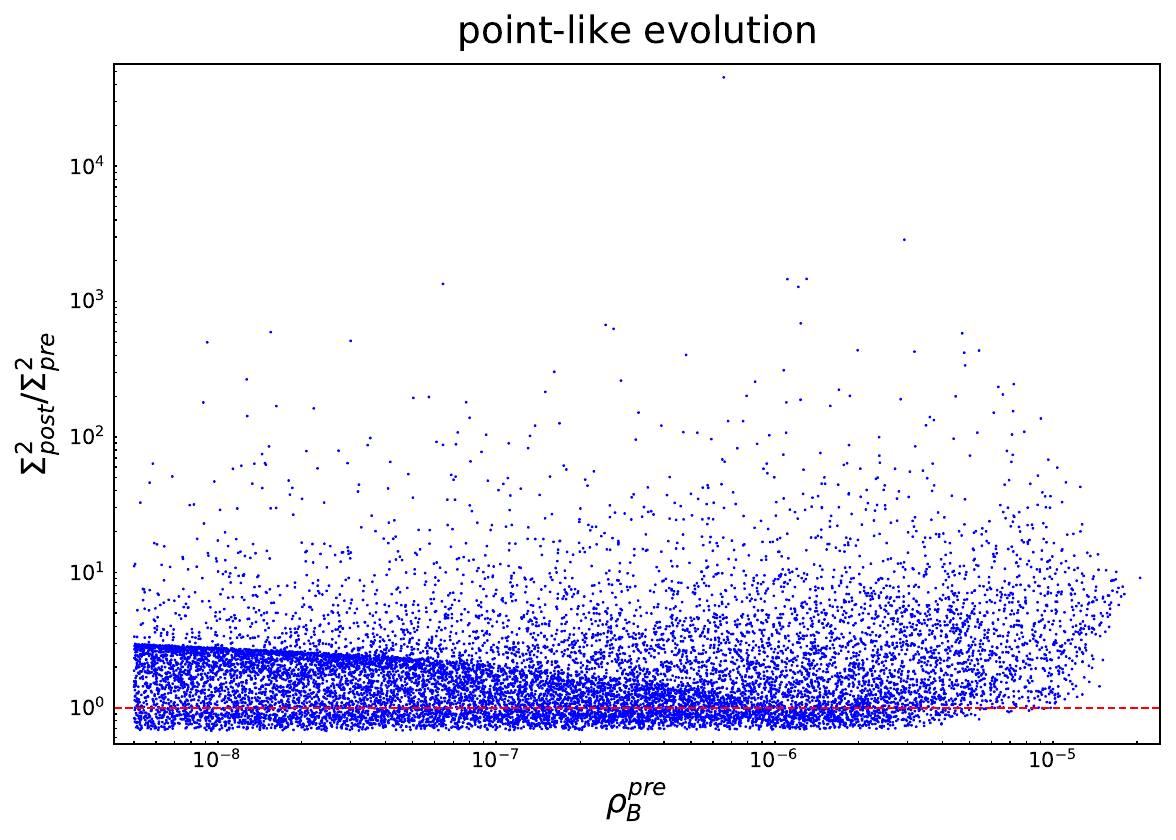} \ \  
    \includegraphics[scale = 0.4]{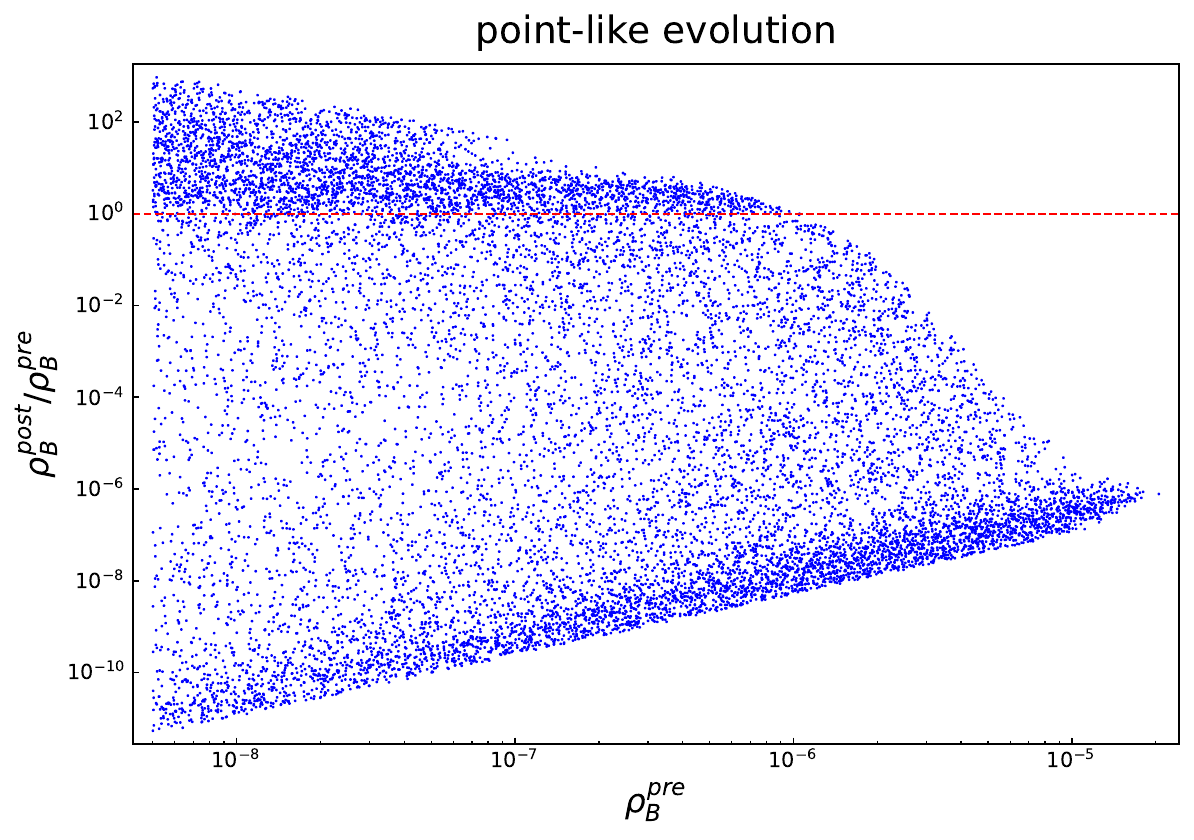}
    \caption{For a classical contracting point-like evolution in the pre-bounce regime with the initial value for the triads set to be $p_{1}(0) = 10^{4}$, $p_{2}(0)= p_{3}(0) = 10^{5}$, while randomizing $c_{2}(0)$ and $c_{3}(0)$ and then specifying the $c_{1}(0) $ by the Hamiltonian constraint within the range that satisfy the classicality condition, i.e., $|\bar \mu_{i} c_{i}|<0.01$, the ratio of post-bounce to pre-bounce values of $\Sigma^2$ and magnetic field energy density versus pre-bounce values of magnetic field energy density are shown in the left and right panels, respectively.}
    \label{fig2}
\end{figure}

Confirming the existence of the bounce, we run simulations for randomized values for $c_{2}$ and $c_{3}$ and fix $c_{1}$ from the Hamiltonian constraint while applying the classicality condition without choosing either point-like or cigar-like initial conditions. We find from 10000 simulations that $93$\% of simulations tend to be cigar-like and only $7$\% are point-like. This means that the universe tends to be cigar-like rather than point-like in the presence of a magnetic field. However, one can set the initial conditions to be point-like by brute force. For this reason, we consider cigar-like and point-like initial conditions separately. In the left panel of Fig. \ref{fig1}, we plot the ratio of the post-bounce to the pre-bounce value of anisotropic shear, i.e., $\Sigma^2$, versus the pre-bounce value of magnetic field energy density for a classical contracting cigar-like evolution in the pre-bounce regime. The red dashed line is a reference line where the ratio of post-bounce to pre-bounce value of the anisotropic shear scalar is unity, and each dot denotes one simulation, while each plot contains more than $10000$ simulations. We should point out that since $\Sigma^2$ is not constant in the presence of a magnetic field, this plot measures the deviation of anisotropic shear from the conserved value. From this plot, one can find that the ratio of post-bounce to pre-bounce value of anisotropic shear has a complicated behavior as the pre-bounce value of magnetic field energy density varies, which is expected due to the non-linearity in the expression for anisotropic shear given in Eq. (\ref{shear}). Moreover, the ratio of the post-bounce to the pre-bounce value of the anisotropic shear approximately varies between $10^{-3}$ and $10$, while the pre-bounce value of magnetic field energy density approximately varies between $10^{-18}$ and $10^{-5}$. When the pre-bounce value of magnetic field energy density becomes negligible, i.e., $\rho_B^{\mathrm{pre}}\lessapprox 10^{-17}$, the model is similar to the vacuum Bianchi-I LQC model, and the ratio approaches unity, implying that anisotropic shear is conserved across the bounce, as shown in Ref. \cite{Motaharfar:2023hil} for vacuum spacetime. In the right panel of Fig. \ref{fig1}, we plot the ratio of the post-bounce to the pre-bounce value of magnetic field energy density, which approximately varies between $10^{-7}$ and $10^{9}$, versus the pre-bounce value of magnetic field energy density. In this plot, we observe that the ratio of the post-bounce to the pre-bounce value of magnetic field energy has a larger value for a smaller initial value of magnetic field energy density, and it decreases as the initial value of magnetic field energy density increases. In fact, for $\rho_B^{\mathrm{pre}} \lessapprox 10^{-11}$, all considered initial conditions result in a larger ratio of post-bounce to pre-bounce value of magnetic field energy density, meaning that the bounce amplifies the post-bounce value of magnetic field energy density in comparison to its pre-bounce value, while for $\rho_B^{\mathrm{pre}}\gtrapprox 10^{-6}$, it suppresses the post-bounce value of magnetic field energy density in comparison with its pre-bounce value for all considered initial conditions. In other words, a quantum bounce due to the quantum gravitational effect acts as a seesaw mechanism for magnetic field energy density in the classical contracting cigar-like evolution in the pre-bounce regime. We also perform simulations starting from the expanding universe, $c_{2}(0)>0$, $c_{3}(0)>0$, and $c_{1}(0)<0$, while the initial value of triads is the same as the one considered for the contracting evolution, and evolve the dynamical equations back towards the bounce. We find that the seesaw mechanism is time reversible. Furthermore, we also change the value of triads, i.e., the physical volume, and find that the cutoff value for which the post-bounce value of magnetic field energy density gets suppressed in comparison with its pre-bounce value for all simulations is independent of the initial value for triads, or correspondingly, the initial volume.

In the left panel of Fig. \ref{fig2}, we plot the ratio of the post-bounce to the pre-bounce value of anisotropic shear versus the pre-bounce value of magnetic field energy density for a classical contracting point-like evolution in the pre-bounce regime. The ratio approximately varies between $0.5$ and $10^{4}$, while the pre-bounce value of the magnetic field approximately changes between $10^{-8}$ and $10^{-5}$. We should point out that, in this case, one cannot extend the range of magnetic field energy density to an arbitrary small value and still satisfy the classicality condition while all three directional connections are negative, i.e., point-like evolution. This is because as the magnetic field energy density becomes negligible, the universe approaches vacuum Bianchi-I spacetime, where the universe has always a cigar-like evolution. In the right panel of Fig. \ref{fig2}, we plot the ratio of the post-bounce to the pre-bounce value of magnetic field energy density versus the pre-bounce value of magnetic field energy density, which approximately varies between $10^{-10}$ and $10^{2}$. From that plot, one can see that the ratio is smaller than unity for $\rho_B^{\mathrm{pre}}\gtrapprox 10^{-6}$, implying that the post-bounce value of magnetic field density gets suppressed in comparison with its pre-bounce value for all considered initial conditions. However, as the initial value of magnetic field energy density decreases, the ratio increases for the number of simulations, indicating that the magnetic field gets amplified through the bounce for those initial conditions, while the reverse happens for the rest of the simulations. In this case, since we cannot extend the range of the initial magnetic field energy density to an arbitrary small value, there is no cutoff value below which the post-bounce value of the magnetic field energy density amplifies in comparison with its pre-bounce value for all considered initial conditions, as is for cigar-like evolution. Although we do not observe the seesaw mechanism for a classical contracting point-like evolution in the pre-bounce regime, the results indicate that the magnetic field gets suppressed for a magnetic field energy density that is approximately larger than $10^{-6}$.

\subsection{Magnetic field production through anisotropic non-singular quantum bounce in the presence of a massless scalar field}

\begin{figure}
    \centering
    \includegraphics[scale = 0.65]{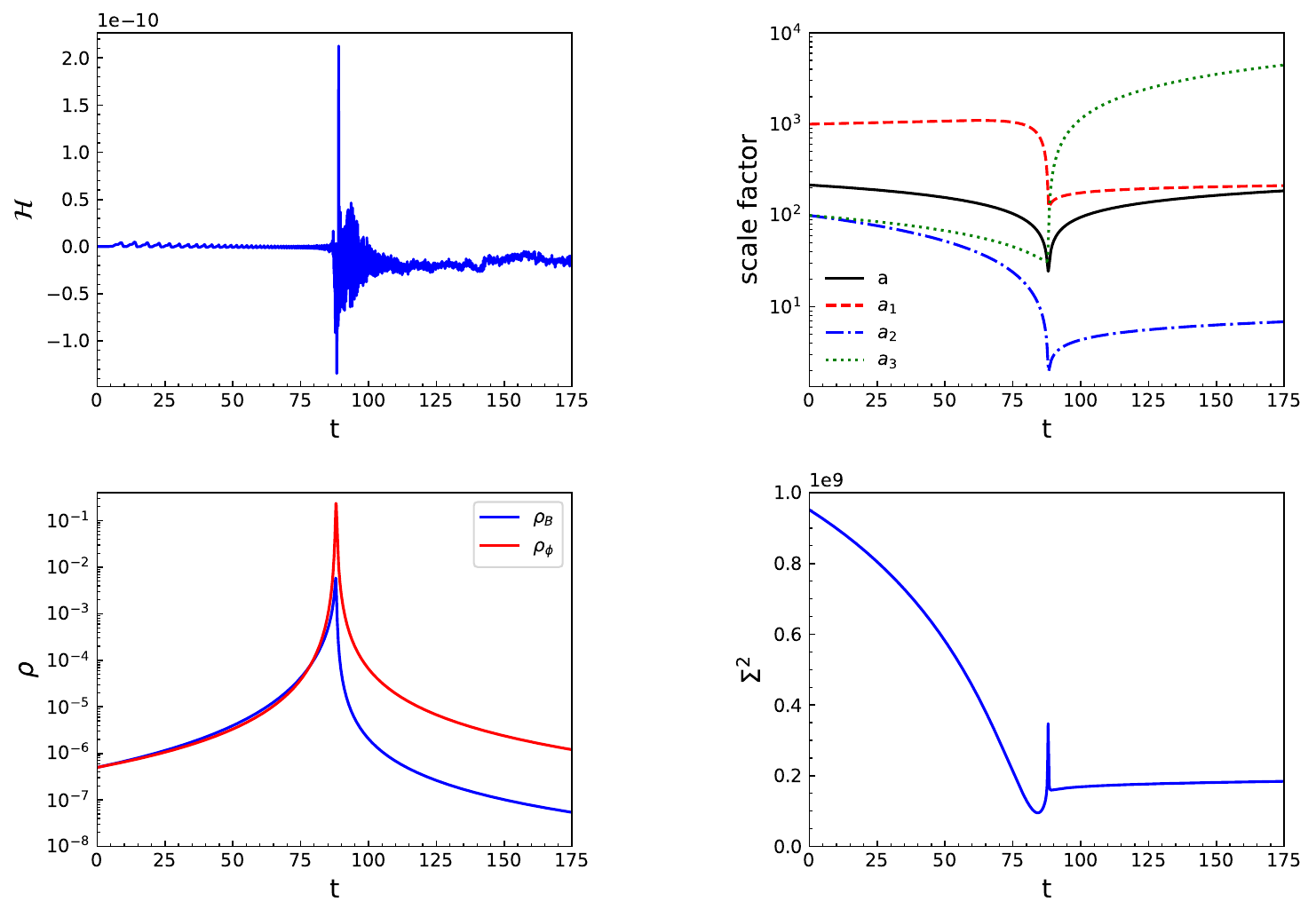}
    \caption{Setting the initial conditions such that the universe starts from a classical contracting cigar-like evolution in the pre-bounce regime, while the initial energy density of massless scalar field is equal to the initial magnetic field energy density, the top left, top right, bottom left, and bottom right panels, respectively, show the time evolution of the Hamiltonian constraint, directional scale factors and the mean scale factor, the magnetic field energy density and the massless scalar field energy density, and $\Sigma^2$. The initial value of triads and connections set as follows: $p_{1}(0) = 10^4$, $p_{2}(0) = p_{3}(0) = 10^5$, $c_{1}(0) = 0.385$, $c_{2}(0) = -0.21$, $c_{3}(0) = -0.13$, and $\rho_{\phi}(0) = \rho_{B} (0) = 5\times 10^{-7}$.}
    \label{cigar-like-massless-1-scale-factor}
\end{figure}

\begin{figure}
    \centering
    \includegraphics[scale = 0.65]{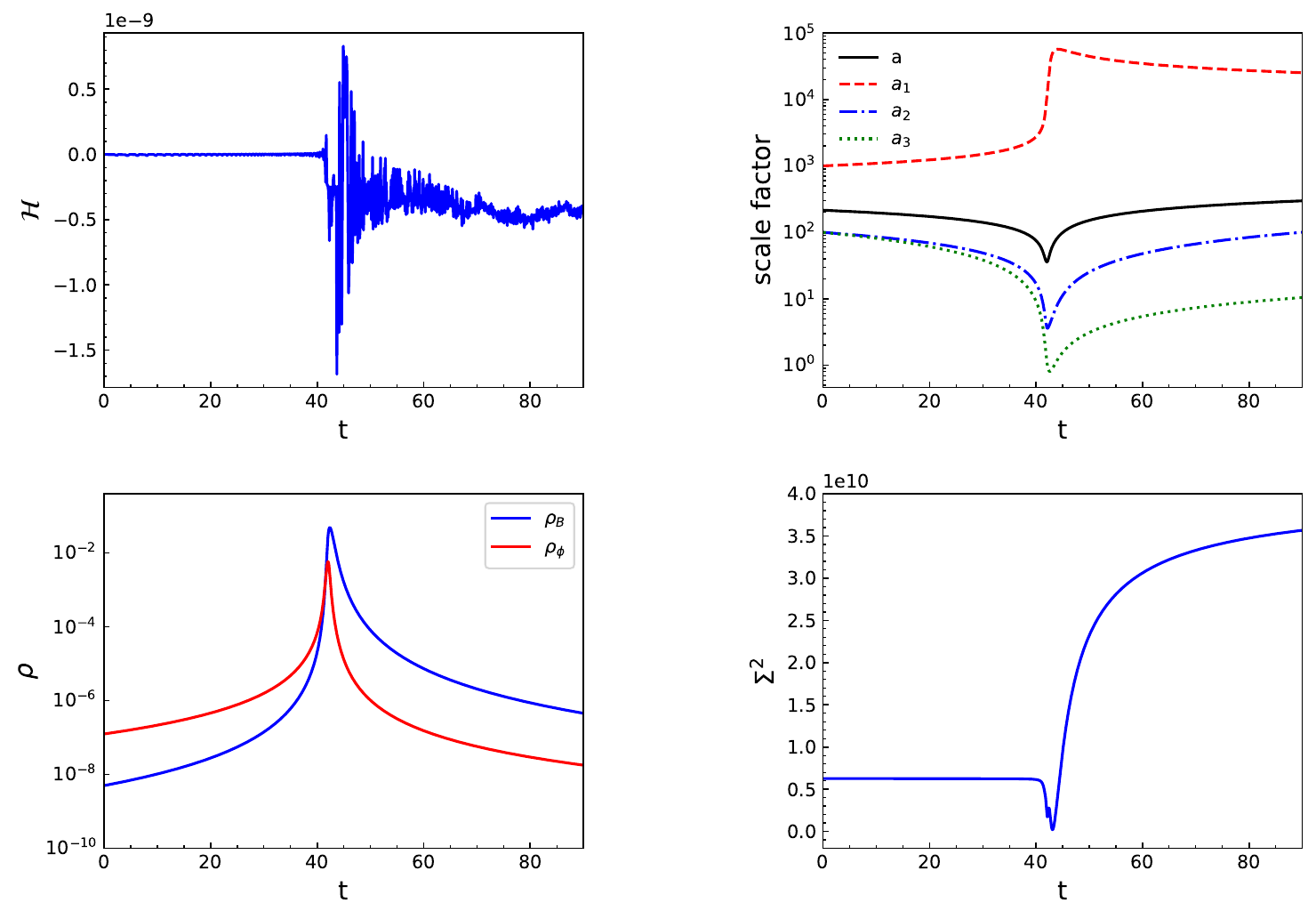}
     \caption{For a classical contracting cigar-like evolution in the pre-bounce regime with the initial values for triads and directional connections fixed at $t=0$ to be $p_{1}(0) = 10^4$, $p_{2}(0) = p_{3}(0) = 10^5$, $c_{1}(0) = 1.81$, $c_{2}(0) = -0.32$, $c_{3}(0) = -0.43$, while the energy density of massless scalar field is dominant, i.e., $\rho_{\phi}(0) = 25 \rho_{B} (0) = 1.25\times 10^{-7}$, the time evolution of the Hamiltonian constraint, directional scale factors and the mean scale factor, the magnetic field energy density and the massless scalar field energy density, and $\Sigma^2$ are shown by top left, top right, bottom left, and bottom right panels, respectively.}
    \label{cigar-like-massless-25-scale-factor}
\end{figure}

Results in the previous section reveal that the quantum bounce acts as a seesaw mechanism for magnetic field energy density starting from a classical contracting cigar-like evolution in the pre-bounce regime, whereby the small or large post-bounce value of magnetic field energy density gets amplified or suppressed across the bounce in comparison with its pre-bounce value. In the case of a classical contracting point-like evolution in the pre-bounce regime, the post-bounce value of magnetic field energy density gets suppressed in comparison with its pre-bounce value for the pre-bounce value of magnetic field energy density approximately larger than $10^{-6}$. However, a universe filled with only magnetic fields cannot describe the early stages of the evolution of the universe. For this, we add a scalar field in this scenario. Ideally, one can consider an inflationary potential; however, it turns out that a massless scalar field suffices to be considered for a couple of reasons. First, in the presence of a positive potential such as inflation, the approach to singularity in the classical FLRW spacetime is kinetic dominated \cite{Foster:1998sk}. This is confirmed in isotropic LQC \cite{Singh:2006im}, as well as for the Bianchi-I model in LQC \cite{Gupt:2013swa}. Further, if one wishes to understand the approach to singularity other than the cigar-like approach, then a massless scalar field is preferred over considering an inflaton potential. The reason is that on approach to singularity, if the scalar field is massless, then its energy density scales as $a^{-6}$, which can balance the growth in anisotropic shear. However, in the presence of a positive potential, the energy density scales as $a^{-6+\epsilon}$, where $\epsilon > 0$, and the singularity structure is always dominated by anisotropic shear. In our analysis, we add a massless scalar field in the presence of a magnetic field and repeat numerical simulations for different ratios of the massless scalar field to the magnetic field energy density at the initial time.

Before moving forward to find out the way the pre-bounce and post-bounce values of anisotropic shear and magnetic field energy density change across the bounce as we vary the ratio of massless scalar field to magnetic field energy density at the initial time, we perform simulations for particular initial conditions to confirm that the singularity is replaced with a quantum bounce in the presence of both a homogeneous magnetic field and a massless scalar field. As an illustrative example, we plot the time evolution of the Hamiltonian constraint, the directional scale factors, the mean scale factor, the magnetic field energy density, the massless scalar field energy density, and $\Sigma^2$ while the initial value of triads and connections is fixed at $t=0$ with $p_{1}(0) = 10^4$, $p_{2}(0) = p_{3}(0) = 10^5$, $c_{1}(0) = 0.385$, $c_{2}(0) = -0.21$, $c_{3}(0) = -0.13$ (classical contracting cigar-like evolution in the pre-bounce regime), and $\rho_{\phi}(0) = \rho_{B} (0) = 5\times 10^{-7}$ in Fig. \ref{cigar-like-massless-1-scale-factor}. As can be seen from the top right, bottom left, and bottom right panels, the mean scale factor $a$, the magnetic field energy density $\rho_{B}$, the massless scalar field energy density $\rho_{\phi}$, and the anisotropic shear $\Sigma^2$ do not diverge in the entire time evolution, signaling that the singularity is resolved in the Bianchi-I LQC model, which includes a homogeneous magnetic field along with a massless scalar field. Moreover, from the top left panel, one can see that the Hamiltonian constraint vanishes throughout the entire time evolution with great accuracy. In addition, from the bottom left and bottom right panels, it is clear that the magnetic field energy density and anisotropic shear have different values before and after the bounce, indicating that the magnetic field can either be suppressed or amplified across the bounce depending on the initial conditions. In Fig. \ref{cigar-like-massless-25-scale-factor}, we plot the time evolution of the Hamiltonian constraint, directional scale factors, the mean scale factor, the magnetic field energy density, the massless scalar field energy density, and $\Sigma^2$ while the initial value of triads and connections is fixed at $t=0$ with $p_{1}(0) = 10^4$, $p_{2}(0) = p_{3}(0) = 10^5$, $c_{1}(0) = 1.81$, $c_{2}(0) = -0.32$, $c_{3}(0) = -0.43$ (classical contracting cigar-like evolution in the pre-bounce regime), and $\rho_{\phi}(0) = 25 \rho_{B} (0) = 1.25\times 10^{-7}$. Again, one can see that all the relevant physical quantities are finite, and the singularity is resolved. Moreover, anisotropic shear scalar and magnetic field energy density have different pre-bounce and post-bounce values in this case. We should note that the singularity is also resolved for point-like initial conditions in the presence of both a homogeneous magnetic field and a massless scalar field as well.

\begin{figure}
    \centering
    \includegraphics[scale=0.4]{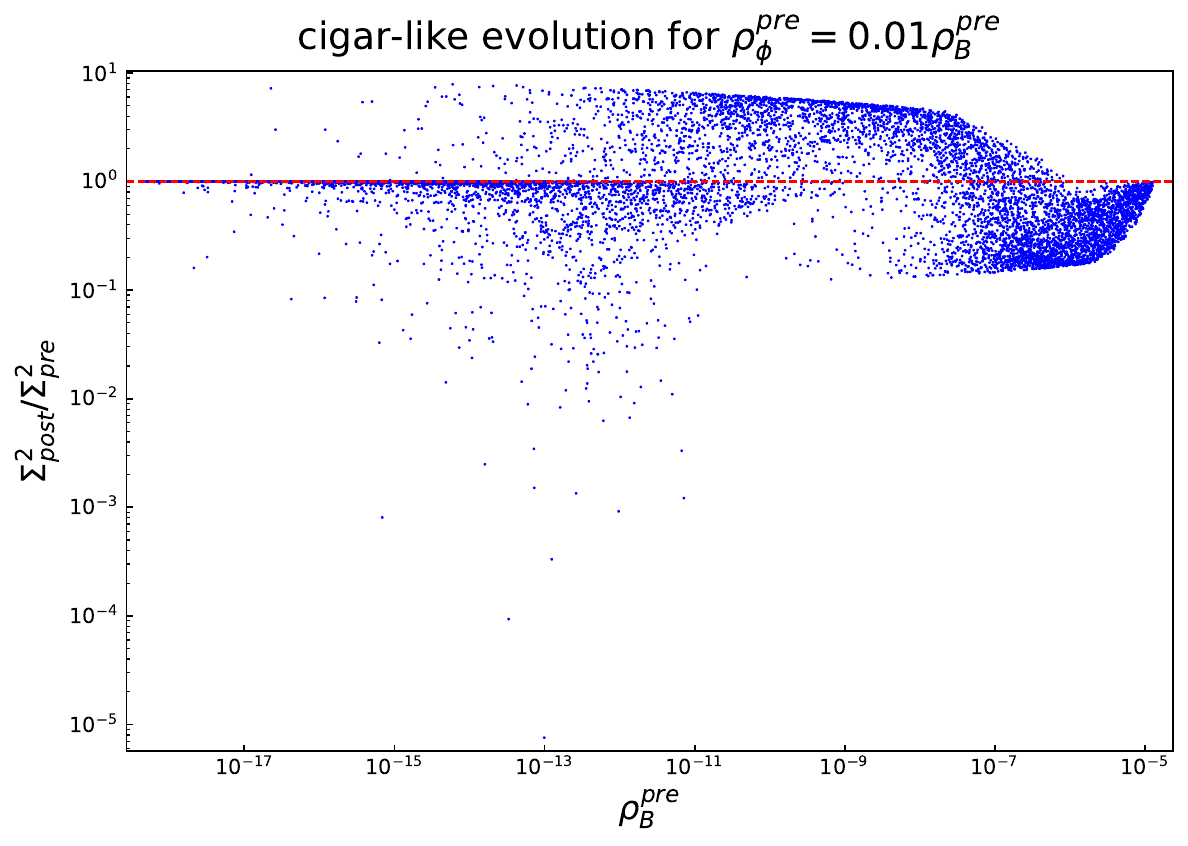} \ \ 
    \includegraphics[scale = 0.4]{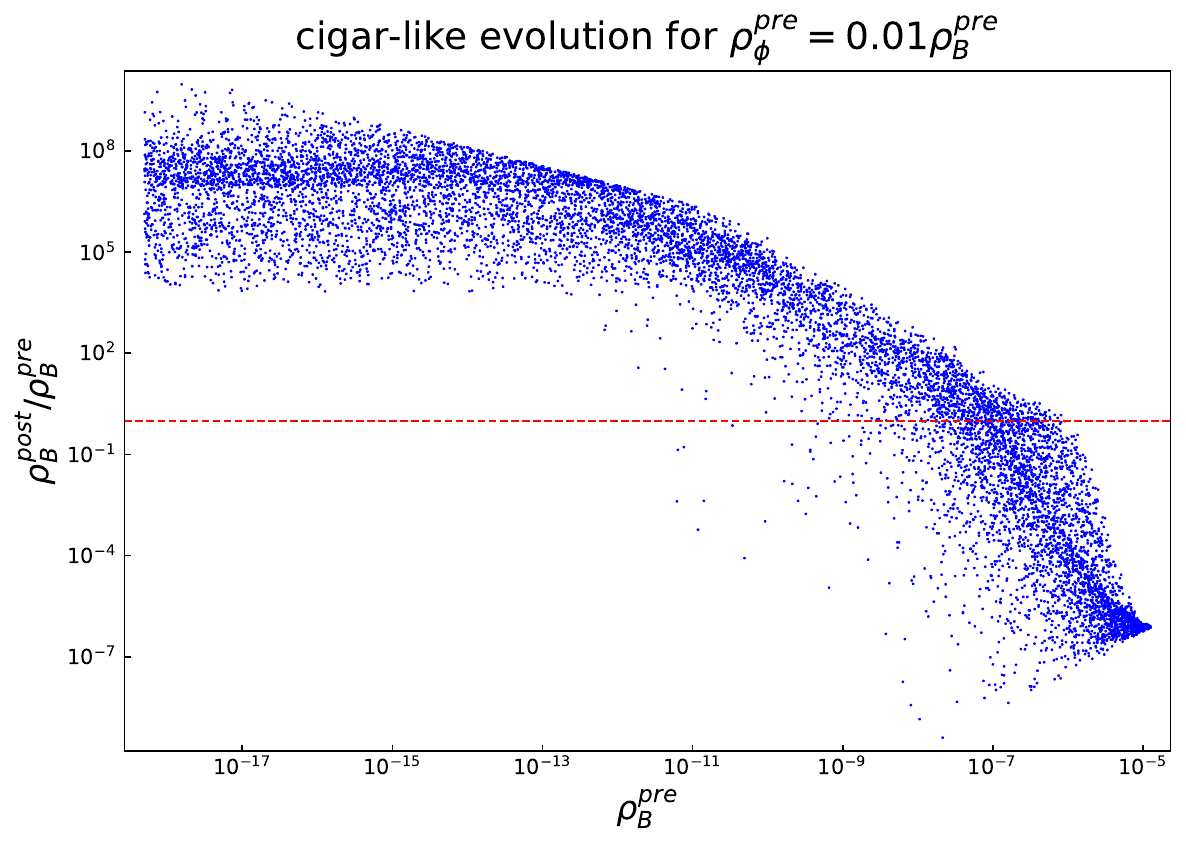}
    \caption{The ratio of post-bounce to pre-bounce value of $\Sigma^2$ versus pre-bounce value of magnetic field energy density (left) and the ratio of post-bounce to pre-bounce value of magnetic field energy density versus pre-bounce value of magnetic field energy density (right) with the initial value of the triads is set as follows: $p_{1}(0) = 10^{4}$, $p_{2}(0)=10^{5}$, $p_{3}(0) = 10^{5}$, and $c_{2}(0)$ and $c_{3}(0)$ are randomized, and $c_{1}(0)$ is then fixed by the Hamiltonian constraint within the range that satisfy the classicality condition, i.e., $|\bar \mu_{i} c_{i}|<0.01$, while the initial energy density of massless scalar field is sub-dominant, i.e., $\rho_{\phi}^{\mathrm{pre}} = 0.01 \rho_B^{\mathrm{pre}}$. The connections are set to start from a classical contracting cigar-like evolution in the pre-bounce regime.}
    \label{fig3}
\end{figure}
\begin{figure}
    \centering
    \includegraphics[scale=0.4]{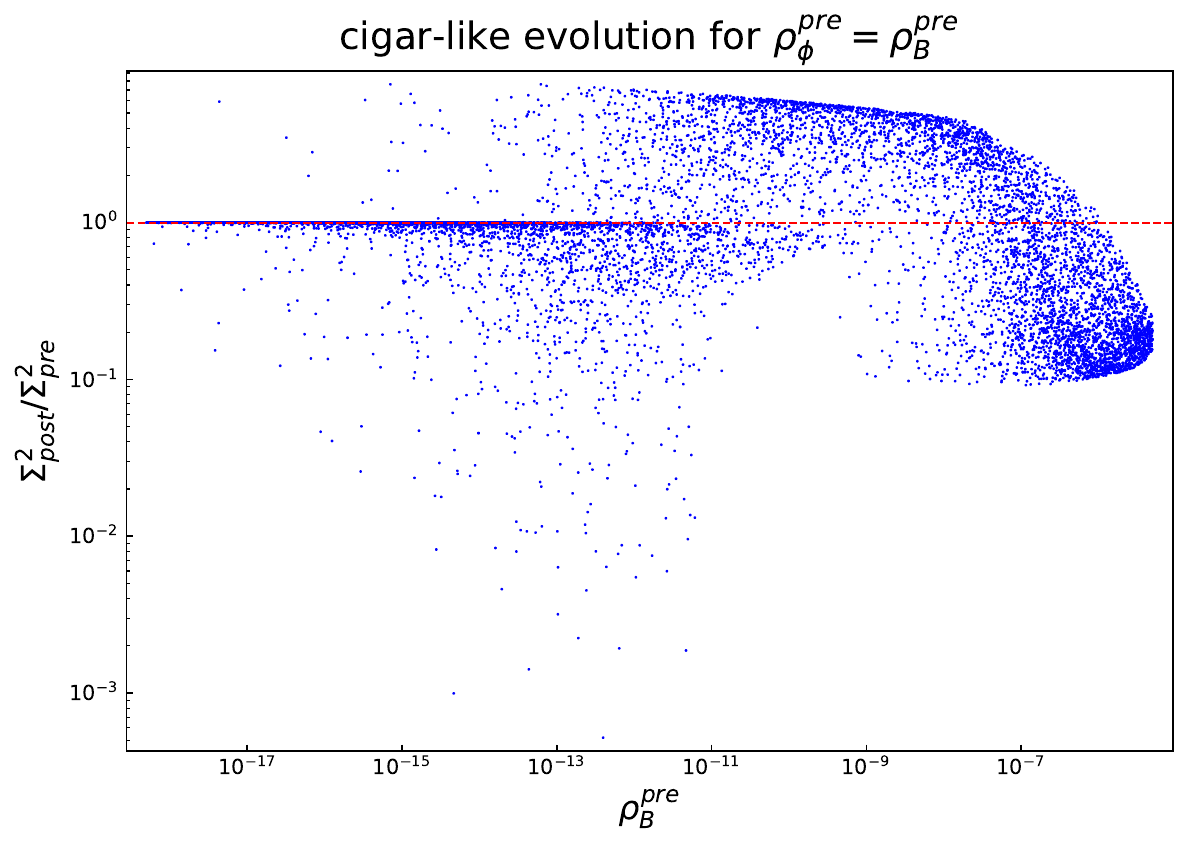} \ \ 
    \includegraphics[scale = 0.4]{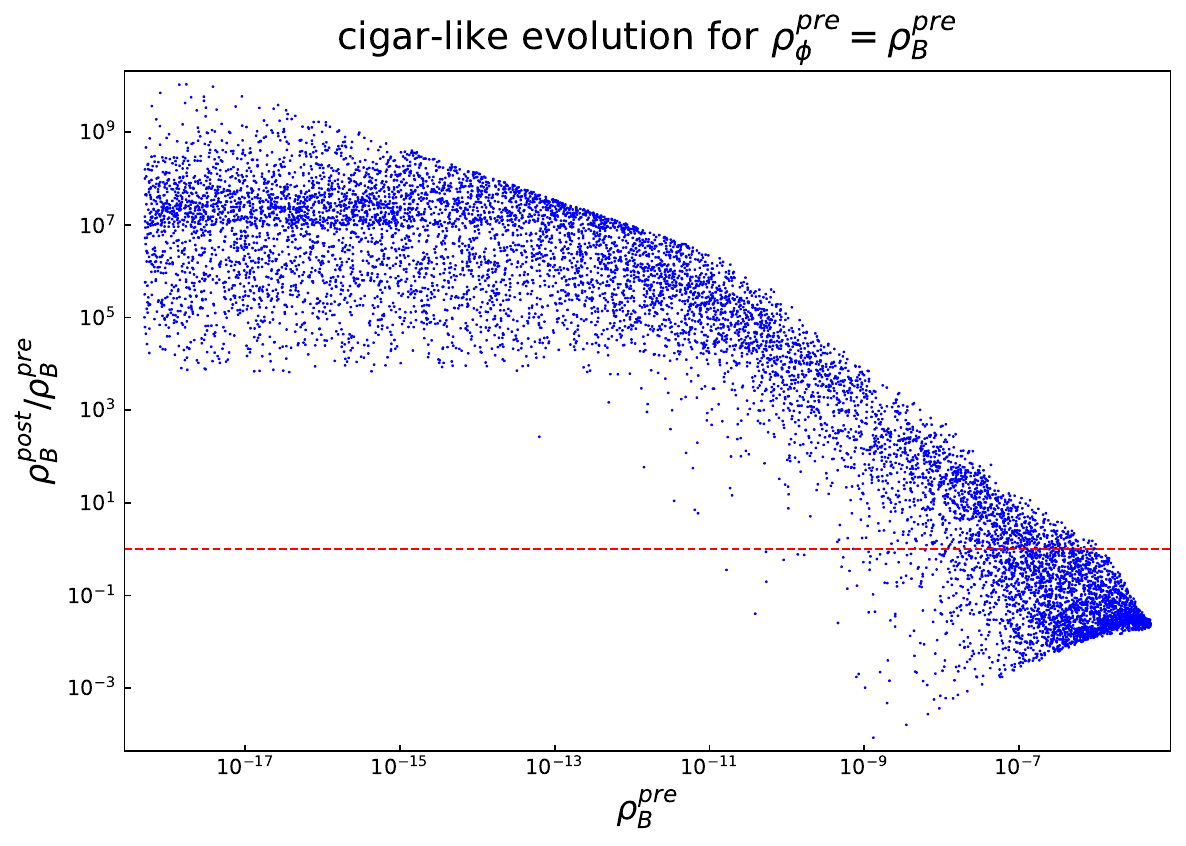}
    \caption{For a classical contracting cigar-like evolution in the pre-bounce regime with initial values for the triads set to be $p_{1}(0) = 10^{4}$, $p_{2}(0)=10^{5}$, $p_{3}(0) = 10^{5}$, while randomizing $c_{2}(0) $ and $c_{3}(0) $, and then specifying $c_{1}(0)$ by the Hamiltonian constraint within the range that they satisfy the classicality condition, i.e., $|\bar \mu_{i} c_{i}|<0.01$, and setting the initial energy density of massless scalar field to be equal to the initial energy density of magnetic field, i.e., $\rho_{\phi}^{\mathrm{pre}} = \rho_B^{\mathrm{pre}}$, the ratio of post-bounce to pre-bounce values of $\Sigma^2$ and magnetic field energy density versus pre-bounce values of magnetic field energy density are shown by left and right panels, respectively.}
    \label{fig4}
\end{figure}
\begin{figure}
    \centering
    \includegraphics[scale=0.4]{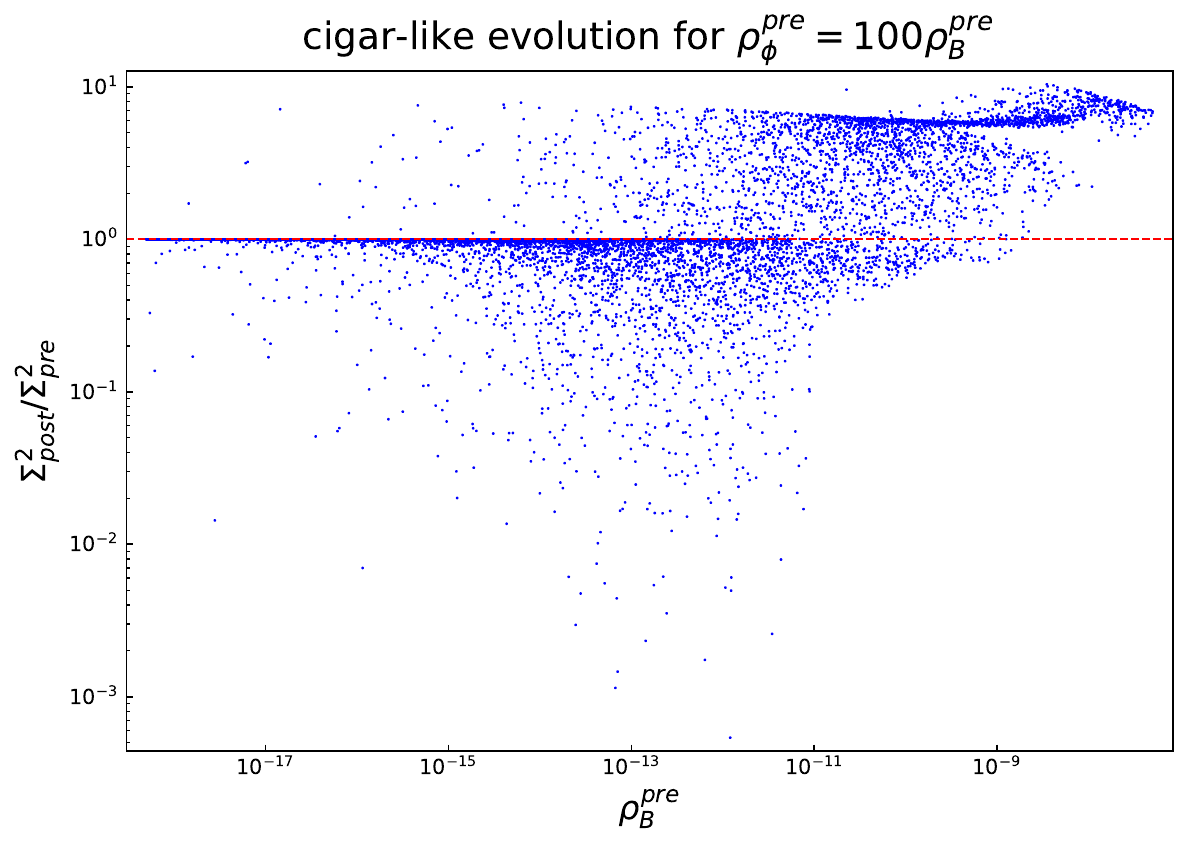}  \ \ 
    \includegraphics[scale = 0.4]{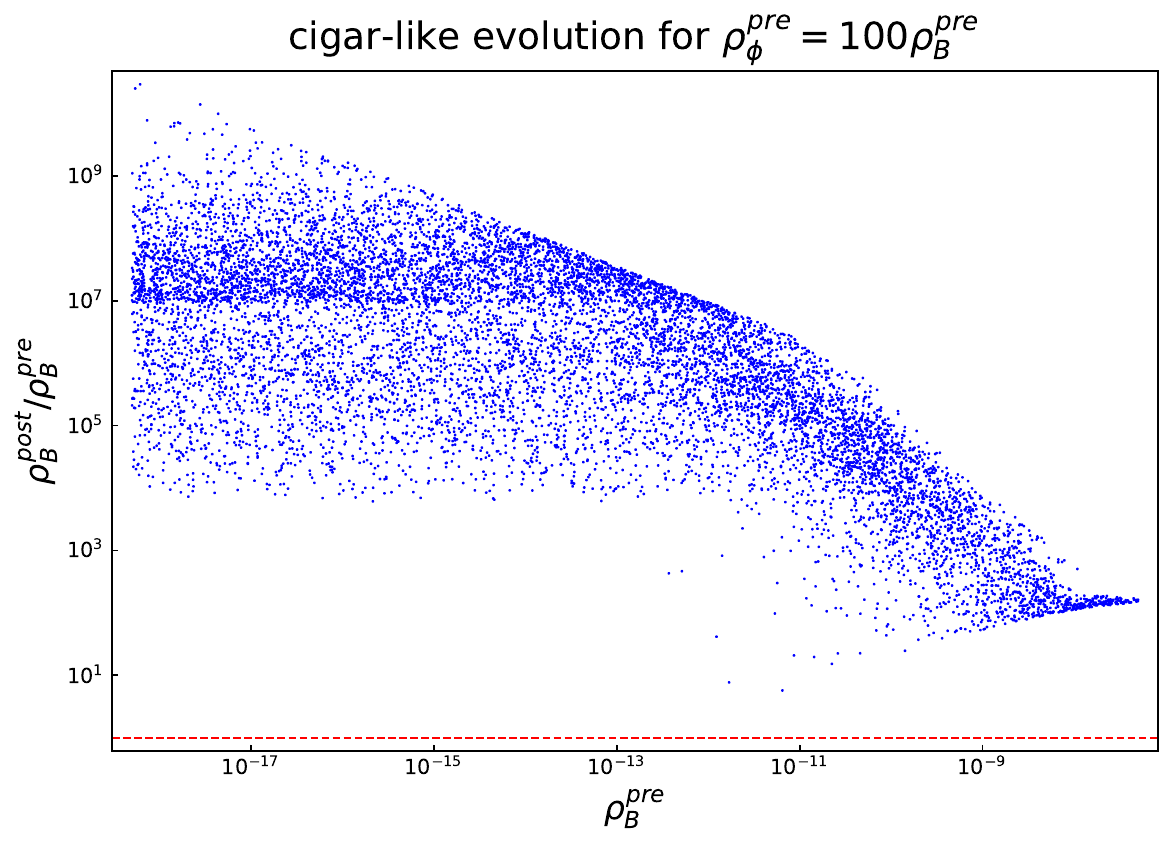}
    \caption{Setting the initial conditions such that the universe starts from a classical contracting cigar-like evolution in the pre-bounce regime, the left and right panels, respectively, show the ratio of post-bounce to pre-bounce values of $\Sigma^2$ and magnetic field energy density versus pre-bounce values of magnetic field energy density in the case that the initial energy density of massless scalar field is dominant. The initial value of the triads is set as follows: $p_{1}(0) = 10^{4}$, $p_{2}(0)=10^{5}$, $p_{3}(0) = 10^{5}$, while $c_{2}(0) $ and $c_{3}(0) $ are randomized, and $c_{1}(0)$ is then fixed by the Hamiltonian constraint within the range that they satisfy the classicality condition, i.e., $|\bar \mu_{i} c_{i}|<0.01$, and $\rho_{\phi}^{\mathrm{pre}} = 100 \rho_B^{\mathrm{pre}}$. }
    \label{fig5}
\end{figure}

\begin{figure}
    \centering
    \includegraphics[scale=0.4]{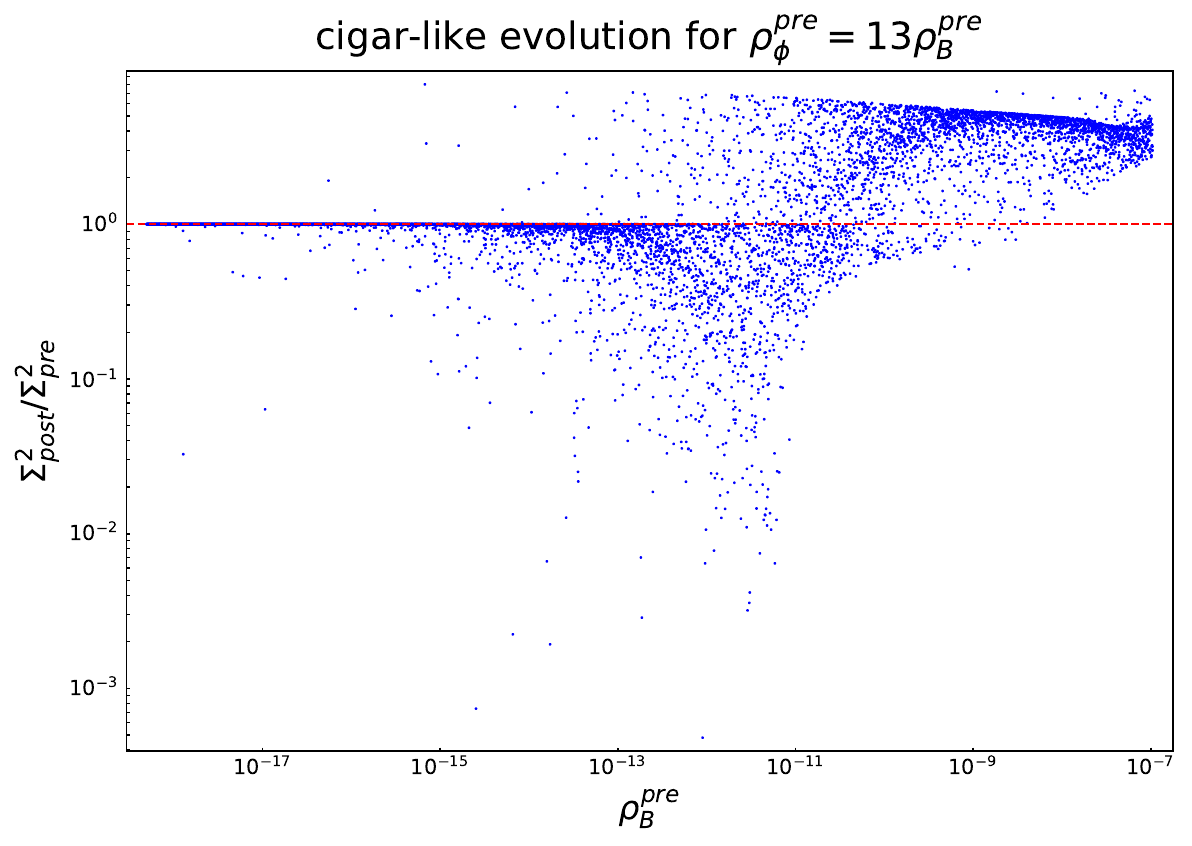}  \ \ 
    \includegraphics[scale = 0.4]{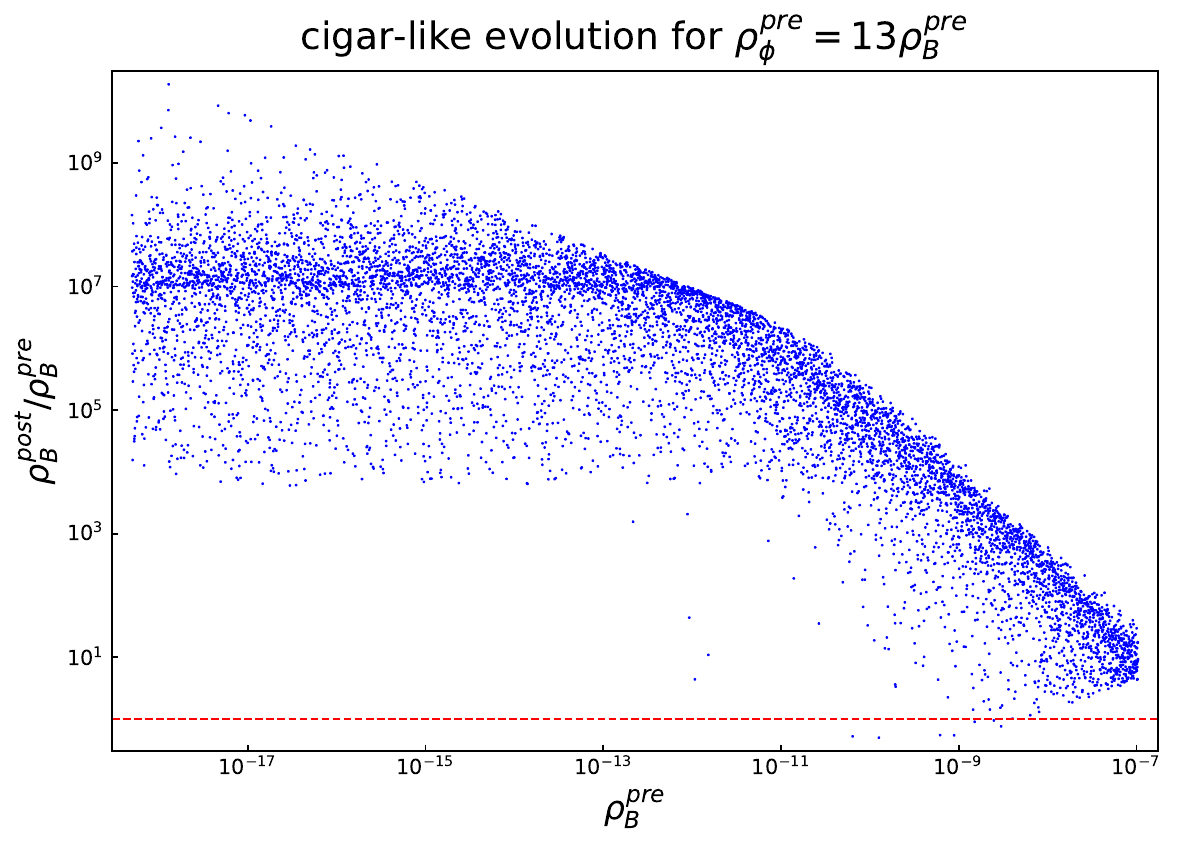}
    \caption{Starting from a classical contracting cigar-like evolution in the pre-bounce regime with the initial value of the triads is set as follows: $p_{1}(0) = 10^{4}$, $p_{2}(0)=10^{5}$, $p_{3}(0) = 10^{5}$, while $c_{2}(0) $ and $c_{3}(0) $ are randomized, and $c_{1}(0)$ is then fixed by the Hamiltonian constraint within the range that they satisfy the classicality condition, i.e., $|\bar \mu_{i} c_{i}|<0.01$, and $\rho_{\phi}^{\mathrm{pre}} = 13 \rho_B^{\mathrm{pre}}$, the ratio of post-bounce to pre-bounce values of $\Sigma^2$ and magnetic field energy density versus pre-bounce values of magnetic field energy density are shown in left and right panels, respectively. }
    \label{massless-13}
\end{figure}

In the left panel of Fig. \ref{fig3}, we plot the ratio of the post-bounce to pre-bounce value of anisotropic shear versus the pre-bounce value of magnetic field energy density for the classical contracting cigar-like evolution in the pre-bounce regime, while the massless scalar field is sub-dominated. In particular, we consider $\rho_{\phi}^{\mathrm{pre}}/\rho_B^{\mathrm{pre}} = 0.01$ at the initial time. One can see that this ratio has almost the same behavior as when a massless scalar field is absent, which is expected since the massless scalar field is sub-dominated. Moreover, from the right panel of Fig. \ref{fig3}, one finds that there is again a seesaw mechanism for the magnetic field whereby the post-bounce value of magnetic field energy density gets suppressed in comparison with its pre-bounce value for $\rho_{B}^{\mathrm{pre}}\gtrapprox 10^{-6}$ and amplified for $\rho_{B}^{\mathrm{pre}}\lessapprox10^{-11}$ as it is the case in the absence of a massless scalar field.
In Fig. \ref{fig4}, we plot the same physical quantities when the energy density of a massless scalar field is equal to the energy density of a magnetic field at the initial time, i.e., $\rho_{\phi}^{\mathrm{pre}} = \rho_B^{\mathrm{pre}}$. From the right panel in Fig. \ref{fig4}, we find that the ratio of the post-bounce to the pre-bounce value of magnetic field energy density approximately varies between $10^{-3}$ and $10^{9}$. In fact, it seems that for a larger number of initial conditions, the post-bounce value of magnetic field energy density tends to get amplified through the bounce in comparison with its pre-bounce value. So it seems that as the massless scalar field gets dominant at the initial time, the ratio of the post-bounce to the pre-bounce value of magnetic field energy density increases for a larger number of simulations. To confirm this point, we finally plot the same physical quantities for the case where the energy density of the massless scalar field is significantly dominated, i.e., $\rho_{\phi}^{\mathrm{pre}}/\rho_B^{\mathrm{pre}} = 100$, at the initial time in Fig. \ref{fig5}. From the right panel, one can observe that for all considered initial conditions, the ratio of the post-bounce to the pre-bounce value of magnetic field energy density is greater than unity, indicating that the post-bounce value of magnetic field energy density gets amplified across the bounce in comparison with its pre-bounce value. In Fig. \ref{massless-13}, by varying the ratio of massless scalar field to magnetic field energy density at the initial time, we find that the post-bounce value of magnetic field energy density gets amplified in comparison with its pre-bounce value for all considered initial conditions when $\rho_{\phi}^{\mathrm{pre}} \gtrapprox 13 \rho_B^{\mathrm{pre}}$ and for triads set to be $p_{1}(0) =10^4$ and $p_{2}(0) = p_{3}(0) = 10^{5}$. Although the lower bound on the initial ratio of massless scalar field to magnetic field energy density can change as we vary the triads, or correspondingly, the volume, the results are unaffected. Thus, we find that the loop quantum bounce leads to magnetic field production if the universe has cigar-like evolution and the massless scalar field energy density significantly dominates.

\begin{figure}
    \centering
    \includegraphics[scale=0.4]{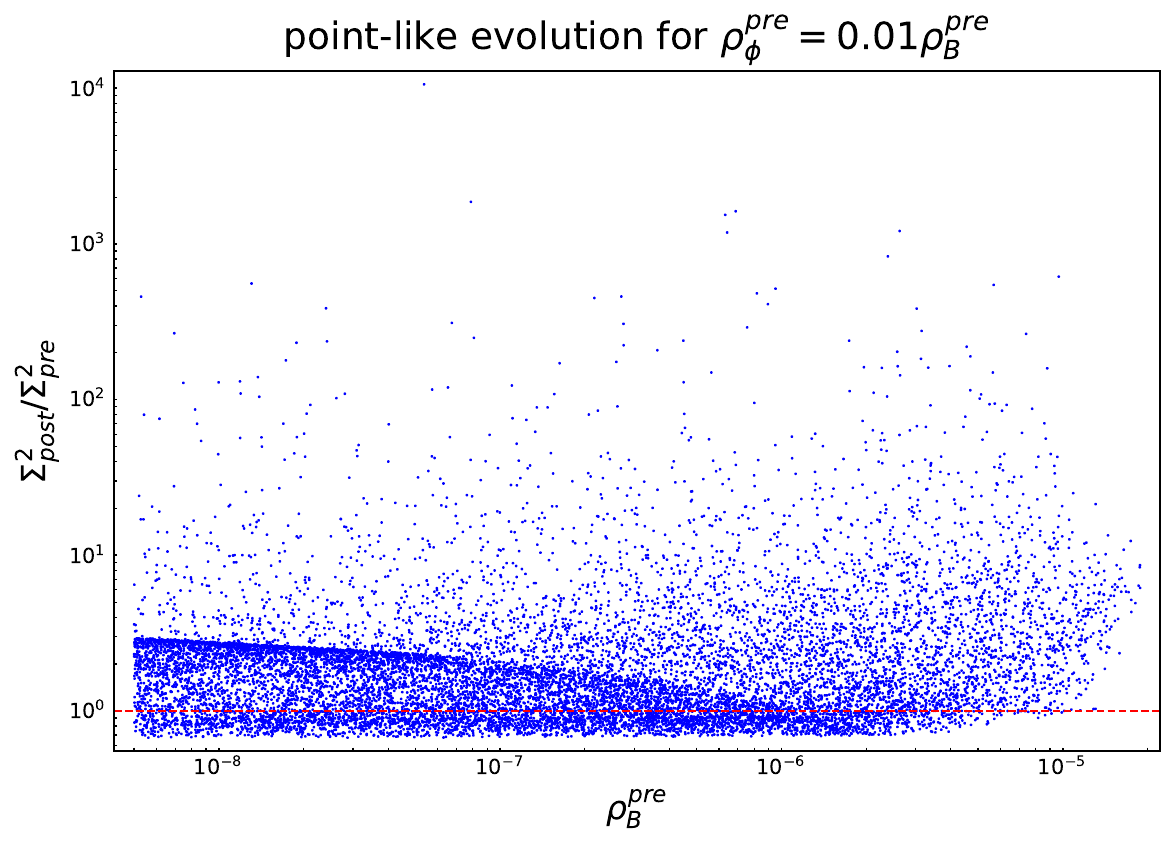} \ \ 
    \includegraphics[scale = 0.4]{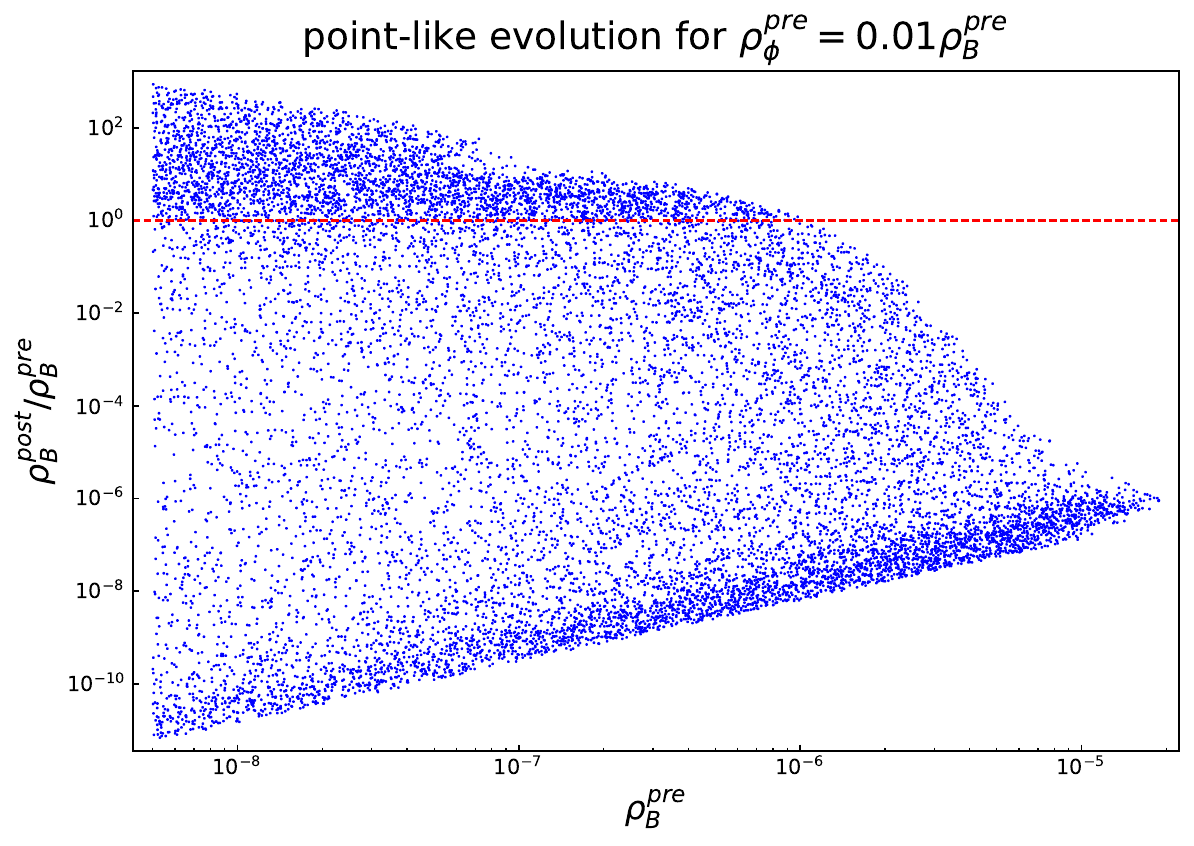}
    \caption{ For a classical contracting point-like evolution in the pre-bounce regime, the left and right panels, respectively, show the ratio of post-bounce to pre-bounce values of $\Sigma^2$ and magnetic field energy density versus pre-bounce values of magnetic field energy density in the case that the initial energy density of a massless scalar field is sub-dominant. The initial value of the triads is set as follows: $p_{1}(0) = 10^{4}$, $p_{2}(0)=10^{5}$, $p_{3}(0) = 10^{5}$, while randomizing $c_{2}(0)$ and $c_{3}(0)$ and then specifying $c_{1}(0)$ by the Hamiltonian constraint within the range that satisfy the classicality condition, i.e., $|\bar \mu_{i} c_{i}|<0.01$, and $\rho_{\phi}^{\mathrm{pre}} = 0.01 \rho_B^{\mathrm{pre}}$. }
    \label{fig6}
\end{figure}

\begin{figure}
    \centering
    \includegraphics[scale=0.4]{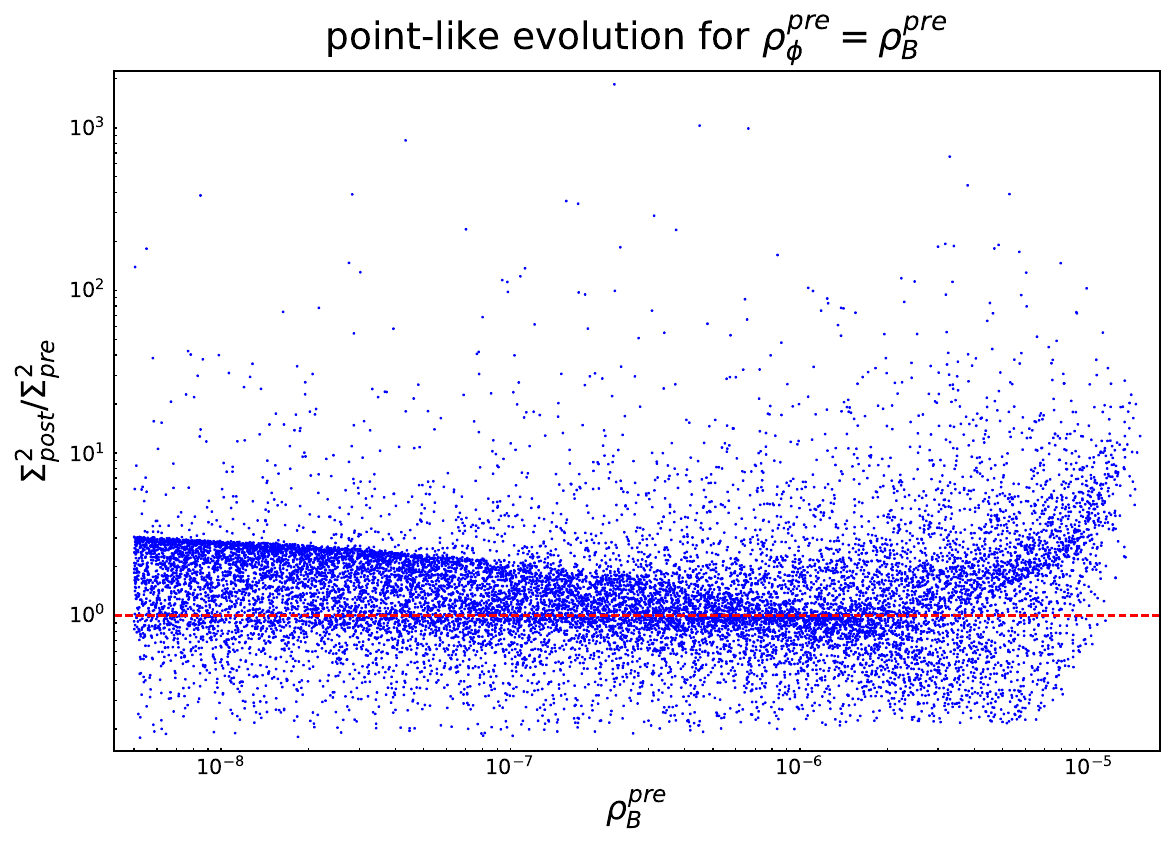} \ \ 
    \includegraphics[scale = 0.4]{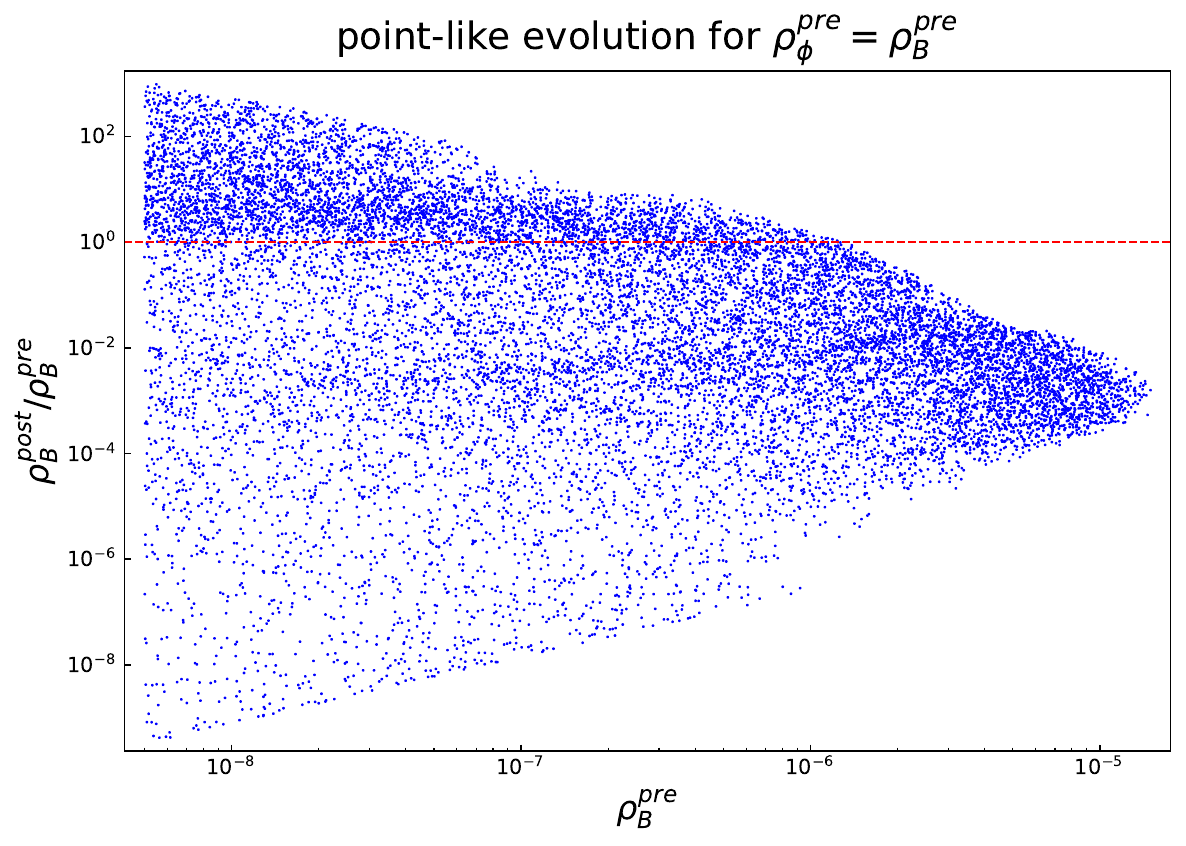}
    \caption{The ratio of post-bounce to pre-bounce value of $\Sigma^2$ versus pre-bounce value of magnetic field energy density (left) and the ratio of post-bounce to pre-bounce value of magnetic field energy density versus pre-bounce value of magnetic field energy density (right) starting from a classical contracting point-like evolution in the pre-bounce regime with the initial value of the triads is set as follows: $p_{1}(0) = 10^{4}$, $p_{2}(0)=10^{5}$, $p_{3}(0) = 10^{5}$, and $c_{2}(0) $ and $c_{3}(0) $ are randomized, and $c_{1}(0)$ is then fixed by the Hamiltonian constraint within the range that they satisfy the classicality condition, i.e., $|\bar \mu_{i} c_{i}|<0.01$, while the initial energy density of massless scalar field is equal to the initial energy density of magnetic field, i.e., $\rho_{\phi}^{\mathrm{pre}} = \rho_B^{\mathrm{pre}}$.}
    \label{fig7}
\end{figure}

\begin{figure}
    \centering
    \includegraphics[scale=0.4]{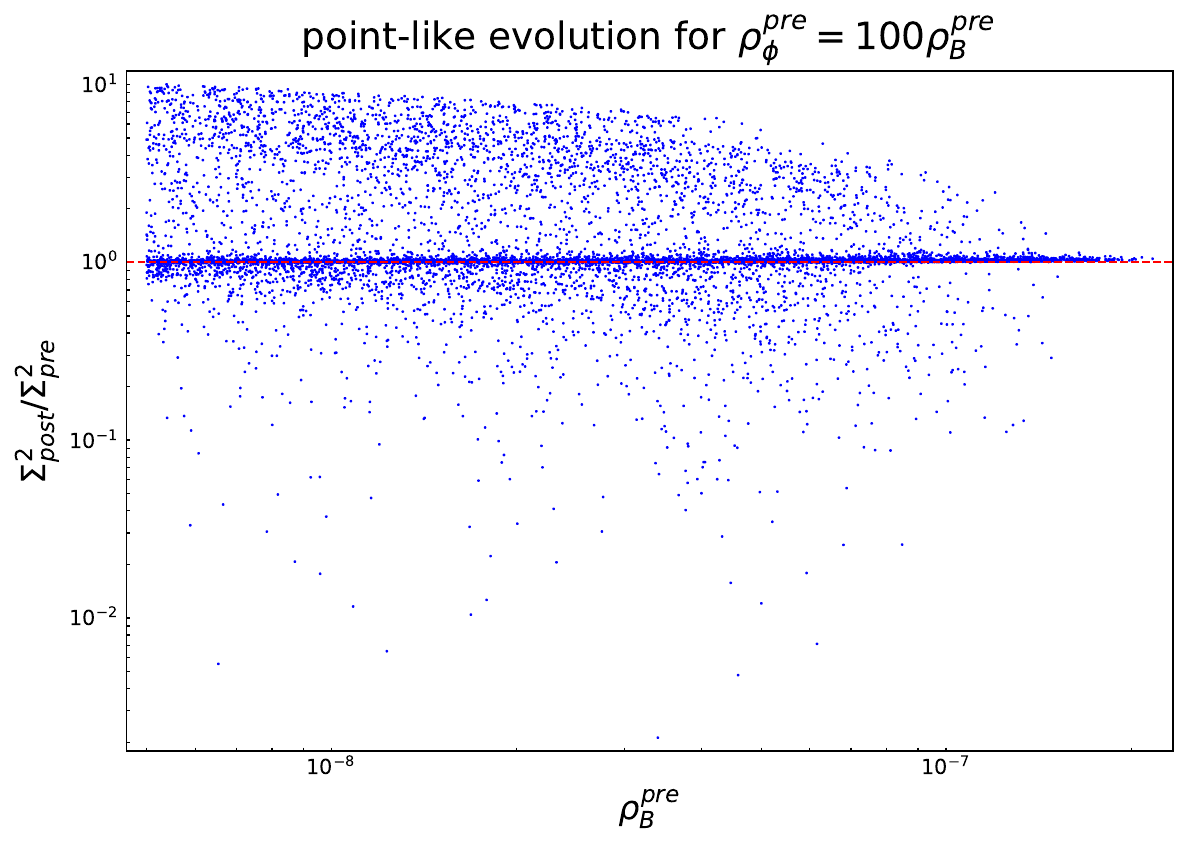} \ \ 
    \includegraphics[scale = 0.4]{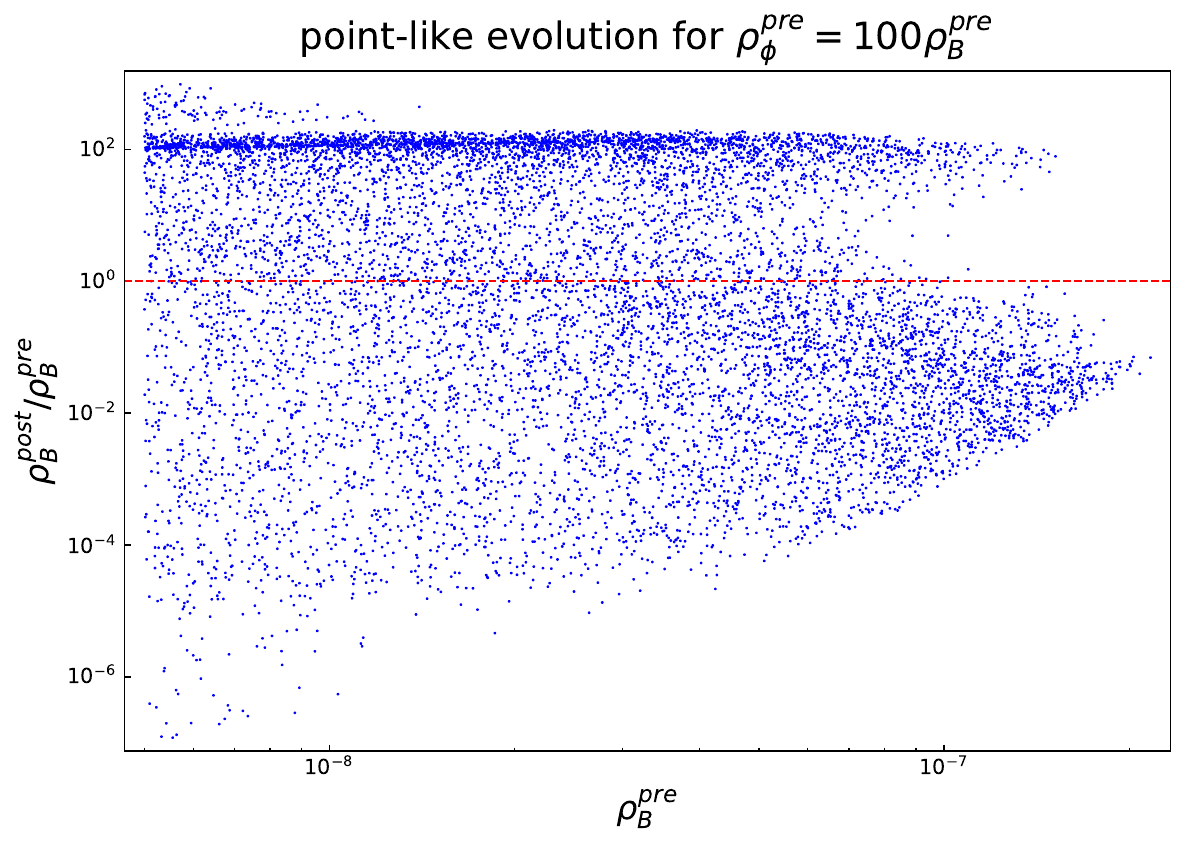}
    \caption{Specifying the initial conditions such that the universe starts from a classical contracting point-like evolution in the pre-bounce regime with $p_{1}(0) = 10^{4}$, $p_{2}(0)=10^{5}$, $p_{3}(0) = 10^{5}$, while randomizing $c_{2}(0) $ and $c_{3}(0) $, and fixing $c_{1}(0)$ by the Hamiltonian constraint within the range that they satisfy the classicality condition, i.e., $|\bar \mu_{i} c_{i}|<0.01$, and setting the initial energy density of massless scalar field to be dominant, i.e., $\rho_{\phi}^{\mathrm{pre}} = 100 \rho_B^{\mathrm{pre}}$, the ratio of post-bounce to pre-bounce values of $\Sigma^2$ and magnetic field energy density versus pre-bounce values of magnetic field energy density are shown by the left and right panels, respectively.}
    \label{fig8}
\end{figure}

In Fig. \ref{fig6}, we plot the ratio of post-bounce to pre-bounce values of anisotropic shear and magnetic field energy density versus pre-bounce values of magnetic field energy density for a classical contracting point-like evolution in the pre-bounce regime, while the massless scalar field is sub-dominated, i.e., $\rho_{\phi}^{\mathrm{pre}}/\rho_B^{\mathrm{pre}} = 0.01$. In this case, the behavior of anisotropic shear and magnetic field energy density is the same as when a massless scalar field is absent. In Fig. \ref{fig7}, we plot the same physical quantities while the massless scalar field and magnetic field have the same order of magnitude, i.e., $\rho_{\phi}^{\mathrm{pre}}/\rho_B^{\mathrm{pre}} = 1$. From the right panel in Fig. \ref{fig7}, one observes that the ratio of the post-bounce to the pre-bounce value of magnetic field energy density approximately varies between $10^{-8}$ and $10^{2}$. So it seems that as the massless scalar field gets dominant, for a larger number of simulations, the post-bounce value of magnetic field energy density gets amplified across the bounce in comparison with its pre-bounce value. However, for a pre-bounce value of magnetic field energy density that is approximately larger than $10^{-6}$, the post-bounce value of magnetic field energy density gets suppressed in comparison with its pre-bounce value for all considered initial conditions, as is the case in the absence of a massless scalar field or a sub-dominated massless scalar field. Finally, in Fig. \ref{fig8}, we plot the ratio of post-bounce to pre-bounce values of anisotropic shear and magnetic field energy density versus the pre-bounce values of magnetic field energy density for the classical contracting point-like evolution in the pre-bounce regime while the massless scalar field is dominated, i.e., $\rho_{\phi}^{\mathrm{pre}}/\rho_B^{\mathrm{pre}} = 100$. From the right panel, one finds that there is no cutoff value below which the post-bounce value of the magnetic field energy density gets suppressed in comparison with its pre-bounce value for all considered initial conditions. That is because the Hamiltonian constraint is not satisfied for a pre-bounce value of magnetic field energy density, which is approximately larger than $10^{-6}$. Hence, the post-bounce value of magnetic field energy density gets either suppressed or amplified in comparison with its pre-bounce value, depending on the initial conditions.

In conclusion, if the universe has a cigar-like evolution in the pre-bounce regime and a massless scalar field is dominant, the magnetic field energy density gets amplified across the bounce, while if the universe has a point-like evolution, the magnetic field energy density can either get suppressed or amplified across the bounce. In other words, the magnetic field's evolution during the quantum bounce is sensitive to initial conditions (cigar-like or point-like evolution), the strength of the magnetic field energy density, and the equation of the state of other matter content in the universe. Hence, if the primordial seed of magnetic fields is produced by a mechanism before the big bang, care should be taken to match the strength of the generated magnetic fields with observational bounds later. In fact, if it turns out that the seed magnetic field produced by pre-big bang magnetogenesis is small and inefficient, it has the potential to get amplified across the quantum bounce to satisfy the observational constraints at later times. On the other hand, if it is significantly large, it will be suppressed and become ineffective to account for the observed value of magnetic fields. 

\section{Summary}\label{section V}

From the size of stars to that of galaxies and clusters, magnetic fields appear to be pervasive throughout the observable universe, based on astronomical observations. It seems that magnetic fields in galaxies and clusters are produced via the amplification of a seed magnetic field of unknown nature, the so-called magnetogenesis mechanism. The success of inflationary cosmology in explaining the generation of matter structure through the amplification of quantum fluctuations in the early universe inspires the idea that magnetic fields may also have a primordial origin. On the other hand, the classical big bang singularity is replaced with a quantum bounce due to quantum gravitational effects in LQC, which opens the possibility for a pre-big bang magnetogenesis mechanism. Hence, it is pertinent to ask if the primordial seed magnetic field is produced by a pre-big bang magnetogenesis mechanism, how it evolves across the bounce, and whether it gets amplified or suppressed through the bounce due to quantum gravitational effects.

To understand the evolution of magnetic fields across a non-singular bounce, we considered a homogeneous magnetic field in the Bianchi-I LQC model. While the anisotropic shear is conserved across the bounce in the classical regime for vacuum spacetime and matter content with vanishing anisotropic stress tensors, it varies during the quantum bounce phase, where quantum gravitational effects dominate. However, the magnetic field is inherently anisotropic, and the anisotropic shear is not preserved across the bounce even in the classical regime. Moreover, the anisotropic shear is closely related to the magnetic field energy density through the anisotropic stress tensor and vice versa. Therefore, it is expected that the magnetic field energy density has a rich behavior as it passes through the bounce due to the non-linear behavior of anisotropic shear and quantum gravitational effects. Thus, it is important to investigate the pre-bounce and post-bounce values of anisotropic shear and magnetic field energy density in the classical regime to explore the evolution of magnetic field across the quantum bounce.

We first performed numerical simulations including a homogeneous magnetic field, initiating with significantly large initial values for triads, correspondingly a large volume, while randomizing the directional connections within the range that satisfy the classicality condition. We set the initial conditions such that the universe starts in a classical regime, contracts, bounces, and expands into a classical expanding regime with different initial values for magnetic field energy density. We find that the universe tends to be more cigar-like than point-like in the presence of a homogeneous magnetic field.
Comparing the pre-bounce and post-bounce values of the magnetic field energy density, we notice an interesting relationship between the pre-bounce and post-bounce values of the magnetic field energy density. We discover that the bounce acts as a seesaw mechanism for the magnetic field such that if the pre-bounce value of magnetic field energy density is approximately less than $10^{-11}$ (in Planck units), the post-bounce value of magnetic field energy density gets amplified in comparison with its pre-bounce value, and if the pre-bounce value of magnetic field energy density is approximately greater than $10^{-6}$, the post-bounce value of magnetic field energy density gets suppressed through the bounce in comparison with its pre-bounce value for all considered initial conditions. By altering the initial value of triads, we demonstrate that the cutoff value, for which the post-bounce value of magnetic field energy density gets suppressed in comparison with its pre-bounce value for all considered initial conditions, remains unaffected by changing the triads. Finally, for a classical point-like universe, which is not common in evolution, we found that the ratio of the post-bounce to the pre-bounce value of the energy density of the magnetic field can either be smaller or greater than unity for the pre-bounce value of the magnetic field energy density, which is approximately less than $10^{-6}$. However, the post-bounce value of magnetic field energy density gets suppressed in comparison with its pre-bounce value for all considered initial conditions when the pre-bounce value of magnetic field energy density is approximately greater than $10^{-6}$, as was the case for the cigar-like universe.

In the next step, including a massless scalar field along with a homogeneous magnetic field, we numerically solved Hamilton's equations to find the evolution of anisotropic shear and magnetic field energy density for different ratios of massless scalar field to magnetic field energy density at the initial time. For the range of initial conditions considered, we found that when the initial energy density of the massless scalar field is approximately greater than 13 times the initial energy density of the magnetic field, the seesaw mechanism disappears, and the magnetic field energy density always gets amplified through the bounce for classical cigar-like universes. More precisely, anisotropic quantum bounce leads to magnetic field production. This also means that the seesaw mechanism is sensitive to the equation of state of the other matter content in the universe. On the other hand, starting from a classical contracting point-like universe, we find that the Hamiltonian constraint is not satisfied for magnetic field energy density approximately larger than $10^{-6}$ as the initial energy density of massless scalar field significantly dominates. Hence, there is no cutoff value for which the post-bounce value of magnetic field energy density gets suppressed in comparison with its pre-bounce value for all considered initial conditions. This means that the post-bounce value of magnetic field energy density gets suppressed or amplified in comparison with its pre-bounce value, depending on the initial conditions. Therefore, a magnetic field's evolution during the quantum bounce depends on the initial conditions (cigar-like or point-like), the strength of the pre-bounce magnetic field energy density, and the equation of the state of other matter content in the universe. 

To conclude, our results shed light on the potential amplification or suppression of a primordial magnetic field across the bounce due to non-perturbative quantum gravitational effects. They indicate that if the primordial seed of the magnetic field is generated before the big bang, one needs to consider the evolution of the magnetic field during the bounce phase to match the strength of the magnetic field to the observed value at a later time. Interestingly, this amplification occurs for a small value of the magnetic field, so if the magnetogenesis mechanism is not efficient enough to amplify the magnetic field, it is possible that it gets amplified during the bounce phase to match the observed value. Conversely, if it turns out that the magnitude of the magnetic field is significantly large, it gets suppressed and becomes inefficient in explaining observational constraints. Our results also hint at the possibility of having a quantum gravitational magnetogenesis due to an anisotropic quantum bounce. In fact, it is known that the bounce is always followed by a super-inflation phase. Since spacetime is anisotropic, the conformal invariance is lost, and a magnetic field can be produced during this super-inflation phase without the need for the inflationary field. Hence, it would be interesting to consider the cosmological perturbations across the bounce in the presence of the magnetic fields to understand that magnetogenesis can occur merely due to quantum gravitational effects, and the predicted value and correlation length are consistent with the observational constraints.

\begin{acknowledgements}
This work is supported by the NSF grant PHY-2110207. All numerical simulations were conducted with high performance computing resources provided by Louisiana State University (http://www.hpc.lsu.edu).
\end{acknowledgements}

\end{document}